\documentclass[aps,prb,twocolumn,superscriptaddress,floatfix]{revtex4}
\usepackage{epsfig,amsmath,amssymb,color}
\begin{document}

\title{Zero-frequency transport properties of one dimensional spin-$\frac{1}{2}$ systems}

\author{F. Heidrich-Meisner}
\email{f.heidrich-meisner@tu-bs.de}
\author{A. Honecker}
\affiliation{Technische Universit\"at Braunschweig, Institut f\"ur Theoretische
  Physik, Mendelssohnstrasse 3, 38106 Braunschweig, Germany}
\author{D. C. Cabra}
\altaffiliation{On leave from Universidad de La Plata and Universidad de Lomas de Zamora, Argentina.}
\affiliation{Ecole Normale Superieure de Lyon, Laboratoire de Physique, All\'{e}e d'Italie, 69264 Lyon, France}
\author{W. Brenig}
\affiliation{Technische Universit\"at Braunschweig, Institut f\"ur Theoretische
  Physik, Mendelssohnstrasse 3, 38106 Braunschweig, Germany}

\date{April 25, 2003}
\begin{abstract}
We report a detailed analysis of the Drude weights for both thermal and
spin transport in  one dimensional spin-$1/2$ systems by means of exact diagonalization and analytic
approaches at finite temperatures. Transport properties are studied first for the integrable $XXZ$
model and second for various nonintegrable systems such as the dimerized chain, the frustrated chain, and the spin ladder.
We compare our results obtained by exact
diagonalization and mean-field theory
with those of  the Bethe ansatz, bosonization and other numerical studies in the case of the anisotropic Heisenberg model
both in the gapless and gapped regime.
In particular, we find indications that the Drude weight for spin transport is finite in the thermodynamic limit
for the isotropic chain.
For the nonintegrable models, a finite-size analysis of the numerical data for the Drude weights is presented covering
the entire parameter space of the dimerized and frustrated chain. We also discuss which conclusions can be drawn from bosonization
regarding the question of whether the Drude weights are finite or not. 
One of our main results is that the Drude weights vanish in the thermodynamic limit for nonintegrable models.

\end{abstract}

\maketitle
\section{Introduction}

Due to their relevance for modeling the magnetic properties of
several quasi-one-dimensional materials, dimerized and frustrated spin-$1/2$
chains and ladders are models of great and current interest. 
Recently, growing experimental evidence has
been found that magnetic excitations can contribute significantly
to the thermal conductivity of various quasi-one and quasi-two-dimensional
materials\cite{ando98,solo00,hess01,kudo01,hess02,solo01,solo03,nakamura91,hofmann03,hess03,sun03}.
Stimulated by these observations, many theoretical activities
have addressed   the issue of heat transport in one-dimensional spin
systems\cite{kane96,saito96,kluemper02,gros02,saito02,hm02,orignac02,saito03a,saito03,louis03,hm03}.
While spin transport has been a topic of numerous theoretical  investigations \cite{shastry90,fye91,scalapino92,bonca94,castella95,zotos96,castella96,zotos97,fabricius98,naef98,narozhny98,fujimoto98,
kirchner99,peres99,zotos99,rosch00,laflorencie01,zotos01,gros02q,rosch02,fujimoto03,long03,louis03,meier03},
the theory of thermal transport is less well understood.\\
One of the key questions is  to understand under which
conditions transport is ballistic, i.e., dissipationless.
The criterion for this is the
existence of a singularity at zero frequency in the real part of
the conductivity. Therefore, one is interested in the integrated
 weight of this singularity - the so called Drude weight - in the
thermodynamic limit. The appearance of a nonzero Drude weight is often
ascribed to the influence of conservation laws on transport
properties\cite{zotos97,rosch00,garst01,fujimoto03,brenig92}. For
example, in the case of the Heisenberg chain the energy-current
operator is conserved\cite{niemeyer71,zotos97} implying a nonzero
thermal Drude weight at all temperatures. Another widely discussed
and related issue is the difference between transport in
integrable models compared
to nonintegrable ones\cite{castella95,narozhny98,gros02q,gros02,fujimoto03,saito03,hm02}. \\
The purpose of the present  paper is  to provide a systematic
study of both the Drude weight for spin and thermal transport at
finite temperatures by means of exact diagonalization on finite systems 
and analytic methods. 
The dependence on exchange coupling
anisotropy, frustration and dimerization will be clarified. 
\\
\indent
We start with an overview of the results of previous and related
works. For the anisotropic Heisenberg chain, it is possible in principle
 to compute thermodynamic quantities exactly with the
Bethe ansatz at arbitrary temperatures. The Drude weight for
thermal transport has been obtained in the antiferromagnetic
regime of this model along this line\cite{kluemper02,kluemper03}.
Numerical studies\cite{gros02,hm02} have provided the thermal Drude weight 
in the antiferromagnetic regime, in the gapped, ferromagnetic phase and 
for the isotropic, ferromagnetic chain.
Here, we extend on such  analysis  
by adding data for the gapless, ferromagnetic regime. \\
\indent
In a first numerical work devoted to the issue of thermal
transport in nonintegrable models, Alvarez and Gros\cite{gros02}
have conjectured that the Drude weight is generically finite in
dimerized and frustrated  spin systems although the energy-current
operator is not conserved in these cases. 
However, we have argued in Refs.\ \onlinecite{hm02,hm03} that this
conclusion cannot be sustained for gapped, frustrated chains if
larger system sizes are included in the finite-size analysis. In this paper, we
extend our parameter study of the thermal Drude weight 
to include dimerized chains and spin ladders also. The main result is
that the numerical data are best interpreted in terms of a {\it
vanishing} thermal Drude weight in  nonintegrable systems.
\\\indent
Recently, however, the thermal Drude weight has  also been computed
by means of analytic approaches yielding a finite Drude weight at
low temperatures\cite{saito03,orignac02}. In Ref.\
\onlinecite{orignac02}, the spin ladder and the dimerized $XY$
model have been studied by mapping to non-interacting models.
While this is exact in the latter case, 
it is an approximation 
in the case of the 
spin ladder. For instance, the influence of  incommensurate umklapp-scattering terms 
on transport properties of massive models is not yet fully understood. 
Bosonization was applied in Ref.\
\onlinecite{saito03} to the cases of the dimerized and the
frustrated chain  leading to the interesting result 
that certain umklapp-scattering terms do not spoil the 
conservation of the energy current.
\\\\\indent
Spin transport in spin-$1/2$ models is equivalent to charge transport of
(spinless) fermions and an  enormous amount of work has been devoted
to this
field\cite{shastry90,fye91,scalapino92,bonca94,castella95,zotos96,castella96,zotos97,fabricius98,naef98,narozhny98,fujimoto98,
kirchner99,peres99,zotos99,rosch00,laflorencie01,gros02q,rosch02,fujimoto03,long03,zotos01,louis03,meier03}.
The situation of spin transport in the integrable model, i.e., the Heisenberg chain,
is different from the case of thermal transport since the
spin-current operator is only conserved in the case of free
fermions ($XY$ model). Nevertheless, the Drude weight is
finite in the gapless regime while numerical\cite{zotos96,naef98} and analytical\cite{peres99} studies have found indications 
that it may  vanish in the gapped cases. The
case of the isotropic chain is still 
the subject of discussion\cite{narozhny98,gros02q,zotos99,fujimoto03,fabricius98,long03}.\\\indent
At present, no final agreement about the results from different Bethe ansatz
computations\cite{zotos99,kluemper,glocke02,fujimoto98} for the
Drude weight has been achieved. 
For example, the Bethe ansatz computations by Zotos\cite{zotos99} predict the Drude weight of the isotropic 
chain to be zero while  Kl\"umper and co-workers\cite{kluemper} have found a finite Drude weight in this case. 
The latter conclusion is in agreement with
 some numerical works
- exact diagonalization\cite{narozhny98,fabricius98} (ED) and Quantum-Monte Carlo
simulations\cite{gros02q} (QMC) - as well as analytic
approaches\cite{fujimoto03}.
Very recently, however, Long et al.\cite{long03}
have developed a novel Lanczos method  for finite temperatures which
has been applied to spin transport in the anisotropic Heisenberg chain with up to
$N=28$ sites. They interpret their numerical results in terms of a 
vanishing Drude weight for the isotropic chain.\\ 
Regarding the temperature dependence of the Drude
weight in the gapless regime contradicting results are reported 
in the literature\cite{zotos99,glocke02,fujimoto03,gros02q,kluemper}.
\\\indent
As far as nonintegrable models are concerned, the effect of
frustration on spin transport has been studied at zero temperature\cite{bonca94},
 while at finite temperatures,
 Ising-like interactions with distance up to 3 have been investigated with
ED\cite{narozhny98,zotos96} and QMC\cite{gros02q}. Fujimoto and
Kawakami\cite{fujimoto03} have discussed transport in
nonintegrable models  in the low-energy limit with bosonization. They find
a nonzero Drude weight provided that particle-hole symmetry is
broken in agreement with Ref.\ \onlinecite{zotos97}. The physical reason is
the finite overlap of the spin-current operator, which is not conserved in
the presence of any umklapp scattering, to the energy-current
operator. \\
In their  work on charge transport in a one-dimensional system\cite{rosch00}, 
Rosch and Andrei have argued that the Drude weight is
 zero in the presence of  
{\em dangerously irrelevant perturbations}
since then all relevant conservation laws are broken.
On the basis of this result, a zero Drude weight is naturally expected in all
nonintegrable and gapless spin models.\\
It will be an additional focus
of the present paper to provide a complete as possible analysis
of spin transport in nonintegrable models such as the dimerized and
frustrated chains as well as  spin ladders. As a main result, the numerical data indicate
a vanishing of the Drude weight in the thermodynamic limit  consistent with Refs.\ \onlinecite{zotos96,rosch00}.
\\\\\indent 
Although it is  beyond the scope of this paper to give
any conclusive explanations of recent transport  experiments on low-dimensional 
spin materials, we will provide a list of
materials for which this work could be of relevance. Whenever
possible we will make connection to these experiments and discuss
implications of our results for their interpretation. First, there
are  surprisingly large magnetic thermal conductivities
observed in the compounds
$\mbox{(Sr,La,Ca)}_{14}\mbox{Cu}_{24}\mbox{O}_{41}$\cite{solo00,hess01,kudo01,hess02}.
Here, a strong magnetic contribution to the thermal conductivity
is believed to be  mediated  by magnetic excitations of spin ladders. Second,
similar findings have been reported for the spin chain materials
$\mbox{SrCu}\mbox{O}_2$, 
$\mbox{Sr}_2\mbox{Cu}\mbox{O}_3$\cite{solo01} and $\mbox{BaCu}_2\mbox{Si}_2\mbox{O}_7$\cite{solo03}. Third, we mention
the anorganic Spin-Peierls material $\mbox{CuGeO}_3$\cite{ando98}
although the physical nature of the heat carrying excitations is still 
under controversial discussion\cite{ando98,hofmann01}. It has been
suggested that the magnetic properties of this material can be
described to some extent by a dimerized and frustrated chain.
Finally,  strong evidence for thermal transport
mediated by magnon excitations in two-dimensional cuprate
compounds has also been
found\cite{nakamura91,hess03,sun03,hofmann03}.
\\\\
The plan of the paper is the following. In Sec.\ \ref{sec:1}, we
introduce the model which is a dimerized and frustrated Heisenberg
chain including an exchange-coupling anisotropy. Next, we
summarize the basic relations  for the transport coefficients and,
particularly,  for the Drude weights as they follow from linear response 
theory in Sec.\  \ref{sec:2}. The definitions for the
current operators will be given at the end of this section. Our
results are presented in Secs.\ \ref{sec:3} and \ref{sec:4}.
First, we discuss transport properties of the anisotropic
Heisenberg chain using mean-field theory and exact
diagonalization. The results will be compared to Bethe
ansatz computations\cite{kluemper02,kluemper03,zotos99},
QMC\cite{gros02q},  and other numerical
studies\cite{gros02,narozhny98,naef98,long03} as well as
analytical works\cite{fujimoto03,laflorencie01}. Second, we consider several
nonintegrable models in Sec.\ \ref{sec:4}.  Using bosonization, we argue that 
vanishing Drude weights are generically expected in massless\cite{rosch00} and massive
nonintegrable spin models.
Numerical data will be
shown for both thermal and spin transport and will be related to
the results of other
groups\cite{gros02,orignac02,saito03,fujimoto03}. Finally, our main
conclusions are summarized  in Sec.\ \ref{sec:5}.

\section{Model}\label{sec:1}
\begin{figure}
\input{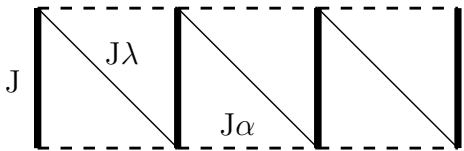}
\caption{Sketch of a frustrated and dimerized $S=1/2$ chain. The limiting cases are:
$\alpha=0,\lambda=1$: the $XXZ$ chain; (ii) $\alpha>0,\lambda=1$: the frustrated chain; (iii)
$\alpha=0,0<\lambda<1$: the dimerized chain; (iv) $\alpha>0,\lambda=0$: the two-leg spin ladder. \label{fig:0}}
\end{figure}
In this work we consider  frustrated and dimerized spin-$1/2$ chains as  sketched in Fig.~\ref{fig:0}.
 The Hamiltonian reads:
\begin{eqnarray} H &=& \sum_{l=1}^{N} h_l(\Delta,\alpha,\lambda) \label{eq:m0} \\
    h_l(\Delta,\alpha,\lambda)&=&      J \lbrack \lambda_l\,h_{l,l+1}(\Delta)
              +\alpha\, h_{l,l+2}(\Delta) \rbrack \label{eq:m1}\\
    h_{l,l+i}(\Delta)&=&     (S_l^+S_{l+i}^-+\mbox{H.c.})/2+{\Delta }S_l^{z}S_{l+i}^{z} .\label{eq:m2}
     \end{eqnarray}
  $h_l(\Delta,\alpha,\lambda)$ from Eq.\  (\ref{eq:m1}) is the local energy density and $N$  the number of sites.
  We set $\lambda_l=1$ if $l$ is even  and $\lambda_l= \lambda$ otherwise. In Eq.\ ({\ref{eq:m2}}), $i=1$
  for nearest- and $i=2$ for next-nearest-neighbor interactions.
 Throughout this work, periodic boundary conditions are used.
 All quantities are measured in units
of the exchange coupling constant $J$ where $J$ is positive.
Mainly, we focus on the following  limiting cases of this  model:
(i) $\alpha=0, \lambda=1$, the $XXZ$ chain;
(ii) $\alpha>0, \lambda=1$, the frustrated chain;
(iii) $\alpha=0, 0<\lambda<1$, the dimerized chain; and
(iv) $\alpha>0, \lambda=0$, the spin ladder.\\
We discuss various values for the anisotropy in the integrable case (i.e., $\alpha=0,\lambda=1$)
including the $XY$ model ($\Delta=0$).
In the case of the  two-leg spin ladder ($\lambda=0$), $\alpha=J_{\parallel}/J_{\perp}$ is equal to the
ratio of the leg coupling $J_{\parallel}$ to the rung coupling $J_{\perp}$.\\
It is instructive to write down the Hamiltonian in terms of fermionic operators. This will later be used
as the starting point for the mean-field theory in Sec.\ \ref{sec:3a}.
By performing a Jordan-Wigner transformation (see, e.g., Ref.\ \onlinecite{mahan}), the 
spin-$1/2$ operators $S_l^{\pm,z}$ can be mapped
onto spinless fermions
\begin{equation}
S^z_l=c_l^{\dagger}c_l^{ }-\frac{1}{2}; \quad S^+_l=e^{i\pi\Phi_l}c_l^{\dagger}.
\label{eq:m4}
\end{equation}
$c_l^{(\dagger)}$ destroys(creates) a fermion on site $l$. The string operator $\Phi_l$ reads
$
\Phi_l=\sum_{i=1}^{l-1}\, c^{\dagger}_ic_i^{ }
$.
For the Hamiltonian from Eq.\  (\ref{eq:m0}), we arrive at (see e.g., Ref.\ \onlinecite{brenig97})
\begin{eqnarray}
H &=& J\sum_l\bigg\lbrack\,\frac{1}{2}\lambda_l\,\big\lbrace (c_l^{\dagger}c_{l+1}^{ }+\mbox{H.c.}) \nonumber \\
        && +\Delta ( \,
        c_l^{\dagger}c_{l}^{ } c_{l+1}^{\dagger}c_{l+1}^{ }-c_l^{\dagger}c_l^{ }+\frac{1}{4})\big\rbrace\nonumber \\
  & & + \alpha \big\lbrace \frac{1}{2}(c_l^{\dagger}c_{l+2}^{ }+\mbox{H.c.})(1-c_{l+1}^{\dagger}c_{l+1}^{ } )\nonumber \\
       &&+ \Delta(
      c_l^{\dagger}c_{l}^{ }c_{l+2}^{\dagger}c_{l+2}^{ }- c_l^{\dagger}c_l^{ } +\frac{1}{4})
      \big\rbrace\bigg\rbrack.\label{eq:m5}
\end{eqnarray}
Note the presence of correlated hopping terms, i.e., 
$c_l^{\dagger}c_{l+2}^{ }c_{l+1}^{\dagger}c_{l+1}^{ }+\mbox{H.c.}$, in Eq.\ (\ref{eq:m5}). Such contributions
are absent in the usual standard extended Hubbard models.
\section{Kubo formulae for $\kappa$ and $\sigma$}\label{sec:2}
In this section we briefly review the underlying equations for the thermal and the
spin conductivity and we define the
energy-current operator $j_{\mathrm{th}}$ and the  spin-current operator $j_{\mathrm{s}}$.


\subsection{Drude weight for heat and spin transport}
The Kubo formulae for the thermal  conductivity $\kappa(\omega)$ and the spin conductivity $\sigma(\omega)$
read\cite{kubo57,luttinger64,mahan,forster} 
\begin{equation}
\kappa[\sigma](\omega) = \frac{[\beta]}{N} \int\limits_0^{\infty} dt\, e^{-i(\omega+i0^+) t}
         \int\limits_0^{\beta} d\tau \langle j_{\mathrm{th[s]}}(-t-i\tau)j_{\mathrm{th[s]}}\rangle. \label{eq:k1}\\
\end{equation}
The expressions in brackets $\lbrack \cdot\rbrack$ refer to spin transport; the index 'th' denotes thermal transport and 's'  spin transport.
$\beta=1/T$ is the inverse temperature while  the brackets $\langle \,.\, \rangle$ denote the thermodynamic
expectation value. The real part of $\kappa(\omega)$ and $\sigma(\omega)$, respectively,
can be decomposed into a $\delta$-function
at $\omega=0$ and a regular part
\begin{equation}
\mathrm{Re}\, \kappa[\sigma] (\omega)= D_{\mathrm{th[s]}}(T) \delta(\omega) +
\mathrm{Re}\,\kappa[\sigma]_{\mathrm{reg}} (\omega)
\label{eq:k2}
\end{equation}
 Here,  we are interested in the  weight  $D_{\mathrm{th[s]}}(T)$ of the $\delta$-type singularity at $\omega=0$.
A nonzero value
of $D_{\mathrm{th[s]}}(T)$ corresponds to a divergent conductivity whereas a vanishing $D_{\mathrm{th[s]}}(T)$
implies a normal  conducting behavior  if there is a contribution from the regular part of the
conductivity at low frequencies or insulating behavior if $\mathrm{Re}\,\kappa[\sigma]_{\mathrm{reg}} (\omega=0)=0$
(see Ref.\ \onlinecite{scalapino92}).\\
In  the remainder of the paper, the dependence on system size of the Drude weight will be explicitly denoted
by $D_{\mathrm{th[s]}}(N,T)$
while $D_{\mathrm{th[s]}}(T)$ refers to the thermodynamic limit $N\to \infty$.\\
\indent
A spectral representation of $\kappa[\sigma](\omega)$ in terms of eigenstates 
$H |\, n\rangle= E_n |\, n\rangle$  yields an expression for both Drude weights\cite{zotos97}
($Z$ is the partition function)
\begin{equation}
D_{\mathrm{th[s]}}^{I}(N,T) =
   \frac{\pi \beta^{t_{\mathrm{th[s]}}}}{Z\,N} \sum_{m, n\atop E_m=E_n}e^{-\beta E_n}
      |\langle m | j_{\mathrm{th[s]}}|n\rangle|^2 .\label{eq:k3}
\end{equation}
Note that the exponent $t_{\mathrm{th[s]}}$ in the prefactor needs to be chosen 
according to
\begin{equation}
t_{\mathrm{th}}=2 \quad \mbox{or} \quad t_{\mathrm{s}}=1,
\label{eq:k3a}
\end{equation} i.e., we expect
$D_{\mathrm{th}}\sim 1/T^2$ and $D_{\mathrm{s}}\sim 1/T$ at  high temperatures.\\
For spin transport, we will numerically analyze  Eq.\ (\ref{eq:k3}) and a second  expression
 which can be derived from linear response theory\cite{shastry90,castella95} 
\begin{equation}
D_{\mathrm{s}}^{II}(N,T) = \frac{\pi}{N}\left\lbrack \langle  - \hat T\rangle -\frac{2}{Z} \sum_{m, n\atop E_m\not=E_n}
 e^{-\beta E_n}\frac{|\langle m |
j_{\mathrm{s}}|n\rangle|^2}{E_n-E_m} \right\rbrack.
\label{eq:k8}
\end{equation}
The operator $\hat T$  is related to the kinetic energy. 
Note that in the derivation of Eq.\ (\ref{eq:k8}), the $f$-sum rule (see Ref.\ \onlinecite{shastry90} and references therein)
is exploited.
It is now instructive to make a connection to   Kohn's formula for the Drude weight $D_{\mathrm{s}}$ 
to illustrate the relation between $D_{\mathrm{s}}^{I}(N,T)$ and $D_{\mathrm{s}}^{II}(N,T)$.
As a by-product, we will obtain  the appropriate definition of $\hat T$.\\
Equation (\ref{eq:k8}) is equivalent to the generalization of  Kohn's formula\cite{kohn64} for finite
temperatures\cite{castella95}
\begin{equation}
D_{\mathrm{s}}^{II}(T)= \frac{\pi}{Z\,N}\sum_n e^{-\beta E_n}
\left. \frac{\partial^2 E_n(\Phi)}{\partial \Phi^2}\right|_{\Phi=0}\label{eq:k10}
.\end{equation}
To arrive at this expression, one considers finite rings of length $N$ being pierced by the  flux $\Phi$,
i.e., a static twist about the $z$ axis. 
The flux-dependent Hamiltonian $H(\Phi)$ with exchange couplings  $J_{r}\not= 0$ being nonzero for $r\leq r_0$
reads\cite{bonca94} 
\begin{eqnarray}
H(\Phi)&=& \sum_{l=1}^{N}\sum_{r=1}^{r_0} J_{r} \big\lbrace \frac{1}{2} ( e^{i (r  \Phi/N )}\, S_{l}^+S_{l+r}^-+\mbox{H.c.})
\nonumber\\
       &&\quad\quad\quad +\Delta S_{l}^z S_{l+r}^z\big\rbrace.
\label{eq:k10a}
\end{eqnarray}
Expanding $H(\Phi)$ up to second order in $\Phi$ yields\cite{shastry90,bonca94}
\begin{equation}
H(\Phi)= H(\Phi=0) + \frac{\Phi}{N} j_{\mathrm{s}} -\frac{\Phi^2}{2\,N^2} \hat T\,.
\label{eq:k10b}
\end{equation}
The expression for the current operator $j_{\mathrm{s}}$ will be given 
in Sec.\ \ref{sec:cop} where we derive it from the equation of continuity.
In terms of spin operators,  $\hat T$ is given by
\begin{equation}
\hat T=  \sum_{l=1}^{N}  \sum_{r=1}^{r_0} r^2\, J_{r}   \,(S_l^+S_{l+r}^-+\mbox{H.c.}).
\label{eq:k9}
\end{equation}
For the models investigated in our paper, we have $r_0\leq 2$. 
In the fermionic picture (see Eq.\ (\ref{eq:m5})), $\hat T$ is the sum of 
the contributions to the kinetic energy from both
the nearest-neighbor and next-nearest-neighbor hopping and also the {\it correlated } hopping terms 
$c_{l+1}^{\dagger}c_{l+1}^{ }c_{l+2}^{\dagger}c_{l}^{ }+\mbox{H.c.}$. The appearance of the extra factor of $4$
in front of the term proportional to $J_2=\alpha\, J$,
which arises from the factor $r^2$ in Eq.\ (\ref{eq:k9}),
can be understood by considering the limit 
of two decoupled chains ($J\alpha=\mbox{const}; J\to 0$) of length $N/2$. Rescaling the size-dependent 
prefactors in Eq.\ (\ref{eq:k10b})  on $N/2$ cancels the factor of $4$ and results in twice the Hamiltonian of a
single chain with 
$N/2$ sites.\\
By applying second-order perturbation theory to $H(\Phi)$ it is straightforward to arrive at 
Eq.\ (\ref{eq:k8}) (see, e.g., Ref.\ \onlinecite{shastry90}).\\  
The relation between $D_{\mathrm{s}}^{I}(N,T)$ and $D_{\mathrm{s}}^{II}(N,T)$ has been pointed out
in Refs.\ \onlinecite{zotos97,narozhny98,scalapino92,kirchner99}.
While
\[D_{\mathrm{s}}(T)=\lim_{N\to\infty} D_{\mathrm{s}}^{I}(N,T)=\lim_{N\to\infty} D_{\mathrm{s}}^{II}(N,T)\]
holds for all temperatures $T>0$ in the thermodynamic limit, the two expressions are not equivalent on {\it finite}
systems. For instance, Eq.\ (\ref{eq:k8}) may result in  negative
values on finite systems at $T=0$ (see Refs.\ \onlinecite{fye91,kirchner99} for examples) whereas $D_{\mathrm{s}}^{I}(N,T)\ge 0$ 
is strictly fulfilled by evaluation
of Eq.\ (\ref{eq:k3}) (see, e.g., Ref.\ \onlinecite{zotos97} for details).
Note in this context that the  difference $D_{\mathrm{s}}^{II}(N,T)-D_{\mathrm{s}}^{I}(N,T)$ 
is proportional to the second derivative of the free energy with respect to the flux $\Phi$ in the limit of vanishing momentum.
In general, this quantity  measures the superfluid density\cite{scalapino92,zotos97,kirchner99}  
in more than one spatial dimension.
However, for a one-dimensional model, it vanishes  in the thermodynamic limit\cite{zotos97,kirchner99}. \\
Eq.\ (\ref{eq:k3}) can also be related to Kohn's formula. While the latter one is the sum of the curvature of energy levels,
$D_{\mathrm{s}}^{I}(N,T)$  equals the sum of squares of first derivatives of the energy levels with respect
to the flux $\Phi$\cite{narozhny98}
\begin{equation}
D_{\mathrm{s}}^{I}(N,T)= \frac{\pi}{Z\, N}\sum_n e^{-\beta E_n}
 \left(\left.\frac{\partial E_n(\Phi)}{\partial \Phi}\right|_{\Phi=0}\right)^2\label{eq:k3b}.
\end{equation}
In this paper, Eqs.\ (\ref{eq:k3}) and (\ref{eq:k8}) will be used to compute the Drude weights while
Eqs.\ (\ref{eq:k10}) and (\ref{eq:k3b}) have been given for completeness. 
We will show in Sec.~\ref{sec:3b} that the
difference between $D_{\mathrm{s}}^{I}(N,T)$ and $D_{\mathrm{s}}^{II}(N,T)$
is  negligibly small at sufficiently high temperatures already on finite systems. 
At low temperatures, only $D_{\mathrm{s}}^{II}(N,T)$ yields a good description of the temperature dependence. 
\indent
If the current operator $j_{\mathrm{th[s]}}$  is a
conserved quantity, Eq.\ (\ref{eq:k3}) can be rewritten as
\begin{eqnarray}
D_{\mathrm{th[s]}}^{I}(N,T) &=&
   \frac{\pi \beta^{t_{\mathrm{th[s]}}}}{Z\,N} \sum_{n}e^{-\beta E_n}
      \langle n | j_{\mathrm{th[s]}}^2| n \rangle \label{eq:k4}\\
      &=& \frac{\pi\beta^{t_{\mathrm{th[s]}}}}{N}\langle j_{\mathrm{th[s]}}^2 \rangle\, , \label{eq:k5}
\end{eqnarray}
implying that a {\it static} expectation value has to be computed which is a considerable simplification. 
Analogously,  we obtain 
\begin{equation}
D_{\mathrm{s}}^{II}(N,T) =  \frac{\pi}{N}\langle -\hat T \rangle.
\label{eq:k5b}
\end{equation}
from Eq.\ (\ref{eq:k8}) under the condition $\lbrack H,j_{\mathrm{s}} \rbrack = 0$.
\indent
The main goal of this work is to investigate whether  $D_{\mathrm{th[s]}}$ has a finite value
in the thermodynamic limit
($N\to \infty$). Typically, the  finite-size  dependence is well controlled at high temperatures
$T\gg J$ where all states $|n\rangle$ are occupied with equal probability $e^{-\beta E_n}\to  1$.
In this case, we see from Eq.\ (\ref{eq:k3}) that
\begin{equation}
 \lim_{T\to \infty}\lbrack T^{t_{\mathrm{th[s]}}}\, D_{\mathrm{th[s]}}^{I}(N,T)\rbrack = C_{\mathrm{th[s]}}(N)
 \label{eq:k6}
\end{equation}
with a prefactor $C_{\mathrm{th[s]}}$ 
\begin{equation}
C_{\mathrm{th[s]}}(N)= \frac{\pi }{Z_{\infty}\,N} \sum_{m, n\atop E_m=E_n}
      |\langle m | j_{\mathrm{th[s]}}|n\rangle|^2.
\label{eq:k7}
\end{equation}
In the limit $T\to \infty$, the partition function $Z_{\infty}=2^N$  is equal to the dimension of the Hilbert space.


\subsection{The current operators}\label{sec:cop}
The current operators obey the following equations of continuity
\begin{eqnarray}
\partial_t h_l &=&i \lbrack H, h_l\rbrack= - (j_{\mathrm{th},l+1}-j_{\mathrm{th},l})
\label{eq:c1}\\
\partial_t S_l^z &=&i \lbrack H, S_l^z\rbrack= - (j_{\mathrm{s},l+1}-j_{\mathrm{s},l})
\label{eq:c2}
\end{eqnarray}
where $h_l$ is the local energy density, $S_l^z$  the local
magnetization density and $j_{\mathrm{th[s],l}}$ are the local
current operators.\\
From Eqs.\ (\ref{eq:c1}) and (\ref{eq:c2}), we obtain   solutions
for the current  operators for arbitrarily long-ranged
interactions. If $\lbrack h_{l\pm r},h_l\rbrack\not= 0$ holds only for
$r\leq r_0$, the  energy-current operator
reads\cite{kawasaki63,niemeyer71,hm02}
\begin{equation}
j_{\mathrm{th}}=\sum_{l=1}^N j_{\mathrm{th},l}=i\sum_{l=1}^N
\sum_{n,r=0}^{r_0-1} \lbrack h_{l-r-1}, h_{l+n}\rbrack
\label{eq:c3}.
\end{equation}
 The spin current is given by\cite{shastry90,bonca94}
 \begin{equation}
j_{\mathrm{s}}=\sum_{l=1}^{N} j_{\mathrm{s},l}=i\sum_{l=1}^N
\sum_{n,r=0}^{r_0-1} \lbrack h_{l-r-1}, S^z_{l+n}\rbrack.
\label{eq:c4}
\end{equation}
with $\lbrack h_{l-r},S_l^z\rbrack\not= 0$    for $r\leq r_0$.
In this paper, we restrict ourselves to $r_0\leq 2$. The same result can be obtained from Eq.\ (\ref{eq:k10b}).
\indent
Independent of the model, the local magnetization density is given
by $S_l^z$ whereas we use
 Eqs.\ (\ref{eq:m1}) and (\ref{eq:m2}) as a definition for the local energy density.
For the $XY$ model, both quantities $j_{\mathrm{th}}$ and $j_{\mathrm{s}}$ commute with
the Hamiltonian, i.e., $\lbrack H^{XY},j_{\mathrm{th[s]}}\rbrack =0$. While $\lbrack H^{XXZ},j_{\mathrm{th}}\rbrack =0$
remains valid for arbitrary $\Delta$ and $\alpha=0;\lambda=1$ \cite{niemeyer71,zotos97},
the conservation of $j_{\mathrm{s}}$ is immediately broken once the anisotropy $\Delta$ has a nonzero value.
Dimerization and frustration which spoil the integrability of the $XXZ$ model also lead to
$\lbrack H,j_{\mathrm{th[s]}}\rbrack\not= 0$ (except for $\Delta=\alpha=0$).

\section{$XXZ$ model}
\label{sec:3}
In this section we discuss transport properties of the $XXZ$ model.
Since this model is solvable via the Bethe ansatz, certain thermodynamic
quantities like susceptibility\cite{eggert94,kluemper00} or specific heat\cite{kluemper00,kluemper02}  are accessible with this technique.
 The thermal Drude weight $D_{\mathrm{th}}$ has recently
been computed along this line by Kl\"umper and Sakai\cite{kluemper02,kluemper03} for arbitrary $\Delta\geq 0$. 
For  spin transport, finite-temperature
Bethe ansatz computations have been performed by both Zotos\cite{zotos99} and Kl\"umper and coworkers\cite{kluemper}.
However, these two groups find different results for the temperature dependence which is ascribed to
conceptual problems of computing the Drude weight from  a single microstate at a given
temperature\cite{glocke02,kluempercomm}.  \\
\indent
First, we apply mean-field theory
based on the Jordan-Wigner transformation and compute the Drude weights in the gapless, antiferromagnetic  regime $0\leq \Delta \leq 1$.
This approach yields a  good approximation
to the exact results from the Bethe ansatz (compare Ref.\ \onlinecite{hm02}) for the thermal conductivity
whereas for spin transport, it fails to describe the $\Delta$-dependence of the Drude weight $D_{\mathrm{s}}$.
Second, we present our results from exact diagonalization  focusing mainly on  spin transport.
\subsection{Mean-field theory}
\label{sec:3a}
In this section, we consider an approximate treatment of the  Hamiltonian of the anisotropic chain  within 
mean-field theory. First, this approach will  provide
exact expressions for the Drude weights in  the case of  free fermions ($\Delta=0$) which 
is useful for a check of the numerical implementation. Second, we will see that the 
thermal Drude weight of the $XXZ$ chain can be well described by this approximation for $|\Delta|\leq 1$\cite{hm02}
which is, however, not the case for spin-transport.\\
 Within mean-field theory, the interaction terms in the fermionized Hamiltonian from Eq.\ (\ref{eq:m5})
are simplified  by means of  a Hartree-Fock decomposition.
Since details of the procedure can be found in the literature
(see, e.g., Ref.\ \onlinecite{brenig97}),
we prefer to quote directly the result for the mean-field Hamiltonian ($k$ labels the momentum)
\begin{equation}
H_{\mathrm{MF}} = \sum_k \epsilon_k c_{k}^{\dagger}c_k^{ }
\label{eq:mf1}.
\end{equation}
$H_{\mathrm{MF}}$ is diagonal in momentum space with the tight-binding dispersion
\begin{equation}
 \epsilon_k=- t(\Delta,T)\cos(k).
 \label{eq:mf2}
\end{equation}
The hopping amplitude $t(\Delta,T):=-(1+2A(T)\Delta)$ is modified via the one-particle expectation value
\begin{equation}
A(T)= \langle c^{\dagger}_{l}c_{l+1}\rangle=\frac{1}{\pi}\int_{0}^{\pi}dk\, \cos(k)f(\epsilon_k),
\label{eq:mf3}
\end{equation}
which  needs to be determined self-consistently. Note that  $f(\epsilon)=1/(\mathrm{exp}(\beta\epsilon)+1)$
is the Fermi function.\\
In the mean-field description, the current operators are simply given by
\begin{equation}
j_{\mathrm{th}}^{\mathrm{MF}} = \sum_k \epsilon_k v_k n_k \,; \quad  j_{\mathrm{s}}^{\mathrm{MF}}=\sum_k\,v_k n_k,
\label{eq:mf4}
\end{equation}
with   velocity  $v_k=\partial \epsilon_k/\partial k$ and $n_k=c_k^{\dagger}c_k^{ }$.
The spinon velocity, i.e., $v_0^{\mathrm{MF}}=v_k(T=0)$, follows from Eq.\ (\ref{eq:mf2})
$v_0^{\mathrm{MF}}= 1+2\,A(T=0)\cdot\Delta$ with $A(T=0)=1/\pi$.
From Eq.\ (\ref{eq:k5}), we obtain  the Drude weights
\begin{eqnarray}
D_{\mathrm{th}}^{\mathrm{MF}}(T)&=& \frac{\pi}{N T^2}\sum_k\, (\epsilon_kv_k)^2\,f^2(\epsilon_k)e^{\epsilon_k/T},\label{eq:mf5}\\
D_{\mathrm{s}}^{\mathrm{MF}}(T) &=&\frac{\pi}{N T}\sum_k\, v_k^2\,f^2(\epsilon_k)e^{\epsilon_k/T}.\label{eq:mf6}
\end{eqnarray}
In the case of free fermions, i.e., $\Delta=0$, Eqs.\  (\ref{eq:mf5}) and (\ref{eq:mf6}) are exact and can be evaluated
both on finite systems and in the thermodynamic limit. This provides a useful check for the numerical implementation; 
in particular since
the spin-current operator $j_{\mathrm{s}}$  is independent of $\Delta$ (see Eq.\ (\ref{eq:c4})).
\\\indent
For thermal transport, the mean-field approach provides a good approximation to $D_{\mathrm{th}}(T)$ for
$|\Delta|\leq 1$
as we have shown in Ref.\ \onlinecite{hm02} and as can be seen in the inset of  Fig.\ \ref{fig:1} (a) where we compare
$D_{\mathrm{th}}(T)$ from the mean-field approach (dotted line) with  the Bethe ansatz results
from Ref.\ \onlinecite{kluemper02} (solid line) for $\Delta=0.5$.\\
Regarding the spin transport, we note that the mean-field theory does not  give a qualitatively correct description
of $D_{\mathrm{s}}(\Delta)$.
Equation (\ref{eq:mf6})  results in $D_{\mathrm{s}}^{\mathrm{MF}}(T=0)=v_0^{\mathrm{MF}}(\Delta)$. Basically, the Drude weight increases
with $\Delta$ here because $D_{\mathrm{s}}^{MF}\sim \langle -\hat T \rangle$ essentially measures the increase of the
bandwidth with $\Delta$. This is  in contrast to the exact expression for $D_{\mathrm{s}}(T=0)$\cite{shastry90}
\begin{equation}
D_{\mathrm{s}}(T=0) = \frac{\pi^2}{4}\frac{\sin(\gamma)}{\gamma(\pi-\gamma)}; \quad \Delta=\cos(\gamma)
\label{eq:mf7}
\end{equation}
where $D_{\mathrm{s}}$ decreases with $\Delta$ due to the presence of the interaction terms in Eq.\
(\ref{eq:m5}).
The reason for the failure of the mean-field theory is that the current operator $j_{\mathrm{s}}^{MF}$ is conserved
for $0\leq\Delta \leq 1$ which is not true for the full Hamiltonian (\ref{eq:m0}) and the respective current operator (\ref{eq:c4}).
Note that  the exact Drude weight takes the form $D_{\mathrm{s}}(T=0)\sim K \, v_0$ where $K$ is the Luttinger parameter
and $v_0$ the velocity at $T=0$ (see, e.g., Ref.\ \onlinecite{laflorencie01}).
It is precisely the factor of $K$ that is missing in the mean-field result.
\subsection{Numerical results for the $XXZ$ model}
Now we turn to the discussion of the results from exact diagonalization for
(i) the thermal Drude weight $D_{\mathrm{th}}$
and (ii) the Drude weight $D_{\mathrm{s}}$ for spin transport
of the anisotropic chain. 
Before going into details, let us outline the structure of this  section and point out the main 
results:\\
\indent First, we discuss technical  aspects of the numerical procedure which will also be of relevance
for the nonintegrable models in Sec.\ \ref{sec:4}. In particular, the importance of degenerate states 
and their numerical identification will be emphasized.  \\
\indent
Second, the numerical results for the thermal Drude weight $D_{\mathrm{th}}(N,T)$ for 
$\Delta=\pm0.5,2$ will be presented and compared to the Bethe ansatz\cite{kluemper02,kluemper03}.
In addition, the high-temperature prefactor $C_{\mathrm{th}}$ is computed and 
discussed as a function of $\Delta$.\\
\indent Third, a detailed parameter study of the Drude weight $D_{\mathrm{s}}$ is performed. 
Results for $D_{\mathrm{s}}^{I,II}(N,T)$ are shown for $\Delta=0.5,\pm 1,-2$ as well as for  $\Delta> 1$.
We discuss the temperature dependence of both $D_{\mathrm{s}}^{I}(N,T)$ and $D_{\mathrm{s}}^{II}(N,T)$
and point out  that they are indistinguishable at high temperatures, but exhibit a completely different
finite-size scaling\cite{narozhny98} at low $T$.  
This result, i.e., $D_{\mathrm{s}}^{I}(N,T)=D_{\mathrm{s}}^{II}(N,T)$ for $T$ large enough, 
is of importance since the Drude weight could in principle also be obtained analytically
from Eq.\ (\ref{eq:k3}) or (\ref{eq:k3b}), respectively, which has to our knowledge not been attempted so far.
Further, the influence of logarithmic finite-size corrections at $T=0$ for
the isotropic chain is mentioned\cite{laflorencie01}.
Finally, and most important, we analyze the finite-size scaling of the high-temperature prefactor
$C_{\mathrm{s}}$ and its dependence on $\Delta$. While we can unambiguously confirm 
that $\lim_{N\to \infty}C_{\mathrm{s}}(N)>0$ for $|\Delta|<1$ in agreement with  the results of 
other groups\cite{naef98,narozhny98,gros02q,fujimoto03,zotos99,zotos96,long03}, 
the data for $\Delta=1$ clearly indicate a {\it finite} Drude weight $D_{\mathrm{s}}(T>0)$ as well.
The latter observation is in agreement with Refs.\ \cite{narozhny98,gros02q,fujimoto03,kluemper,fabricius98}
but  contradicts the conclusions of Refs.\ \cite{zotos99,long03}.
\subsubsection{Technical remarks on the numerical procedure} 
We start with several technical remarks on the
numerical procedure  which are relevant for both  integrable and  nonintegrable models. We have performed complete diagonalization for chains with $N\leq 18$
sites exploiting conservation of the $z$-component $S^z_{\mathrm{tot}}=\sum_{l} S_l^z$ of the total spin,
translational invariance, and spin-inversion symmetry in the $S^z=0$ subspaces of systems
with even $N$. The latter symmetry is respected by the energy-current operator $j_{\mathrm{th}}$ but not by the
spin-current operator $j_{\mathrm{s}}$. The  dimensions of the largest subspaces
for a  given momentum $k$
are  $\approx 2400$ for $S^z_{\mathrm{tot}}=1$ and $\approx 2700$ for $S^z_{\mathrm{tot}}=0$ at $N=18$.
In the latter
case, the dimension is  almost reduced by a factor of 2 by spin-inversion symmetry for the
subspaces with odd and even sign under this symmetry.\\
\indent
Another important aspect is the identification of degenerate states (i.e., $E_n=E_m$) in subspaces
labeled by $S^z_{\mathrm{tot}}$ and momentum $k$. This is necessary both in the  evaluation of
Eqs.\  (\ref{eq:k3}) and (\ref{eq:k8}) but becomes irrelevant if the respective
current operator is conserved, leading to  the simpler expression,
Eq.\  (\ref{eq:k4}). The latter   is possible for thermal transport in the $XXZ$ model. \\
For spin transport, however, we have $\lbrack
H,j_{\mathrm{s}}\rbrack \not= 0$ for $\Delta \not= 0$. The
(integrated) distribution of level spacings $\Delta E_n$ is shown
in  Fig.\  \ref{fig:2}. There, the number $I(\epsilon)$ of level
spacings $\Delta E_n=E_{n+1}-E_{n}, E_n<E_{n+1}$ of adjacent
energy levels being smaller than a given value of
$\epsilon$ is plotted versus $\epsilon$ for  $\Delta=0.5$
\begin{equation}
I(\epsilon)=\sum_{(S_{\mathrm{tot}}^z,k)}\sum_{\Delta E_n<\epsilon}\, 1\, .
\label{eq:int-1}
\end{equation}
It is sufficient to analyze  all subspaces with given $S^z_{\mathrm{tot}}$ and momentum $k$
separately and sum over all subspaces thereafter (indicated by the first sum in Eq.\ (\ref{eq:int-1})).\\
The spectrum  displays some characteristic
features: first, the value of $I(\epsilon)$ for $\epsilon \to \infty$ equals the dimension of the Hilbert space $2^N$ minus the number
of subspaces ($S_{\mathrm{tot}}^z,k$). Second, a large fraction of degenerate states is  present
and third, the integrated distribution of  level spacings is
constant for $10^{-9}J\lesssim \epsilon\lesssim 10^{-6}J$
for the system sizes investigated here. This suggests 
that adjacent energy levels  are typically separated by $\Delta E_n \lesssim 10^{-9}J$
if they are degenerate. \\
This  separation allows for an identification of degenerate states  by imposing
the following criterion in the numerical analysis:
energy levels with  $|E_{n+1}-E_n| < \epsilon_{\mathrm{cut}}=10^{-8}J$ are
degenerate. By evaluation of Eq.\ (\ref{eq:k3})
for the thermal Drude weight we find agreement with Eq.\ (\ref{eq:k4})
and the Bethe ansatz\cite{kluemper02}
for this choice of $\epsilon_{\mathrm{cut}}$ but significant deviations at temperatures $T\sim J$
if larger or lower values for the cutoff energy  are used.


\begin{figure}[t]
\centerline{\epsfig{figure=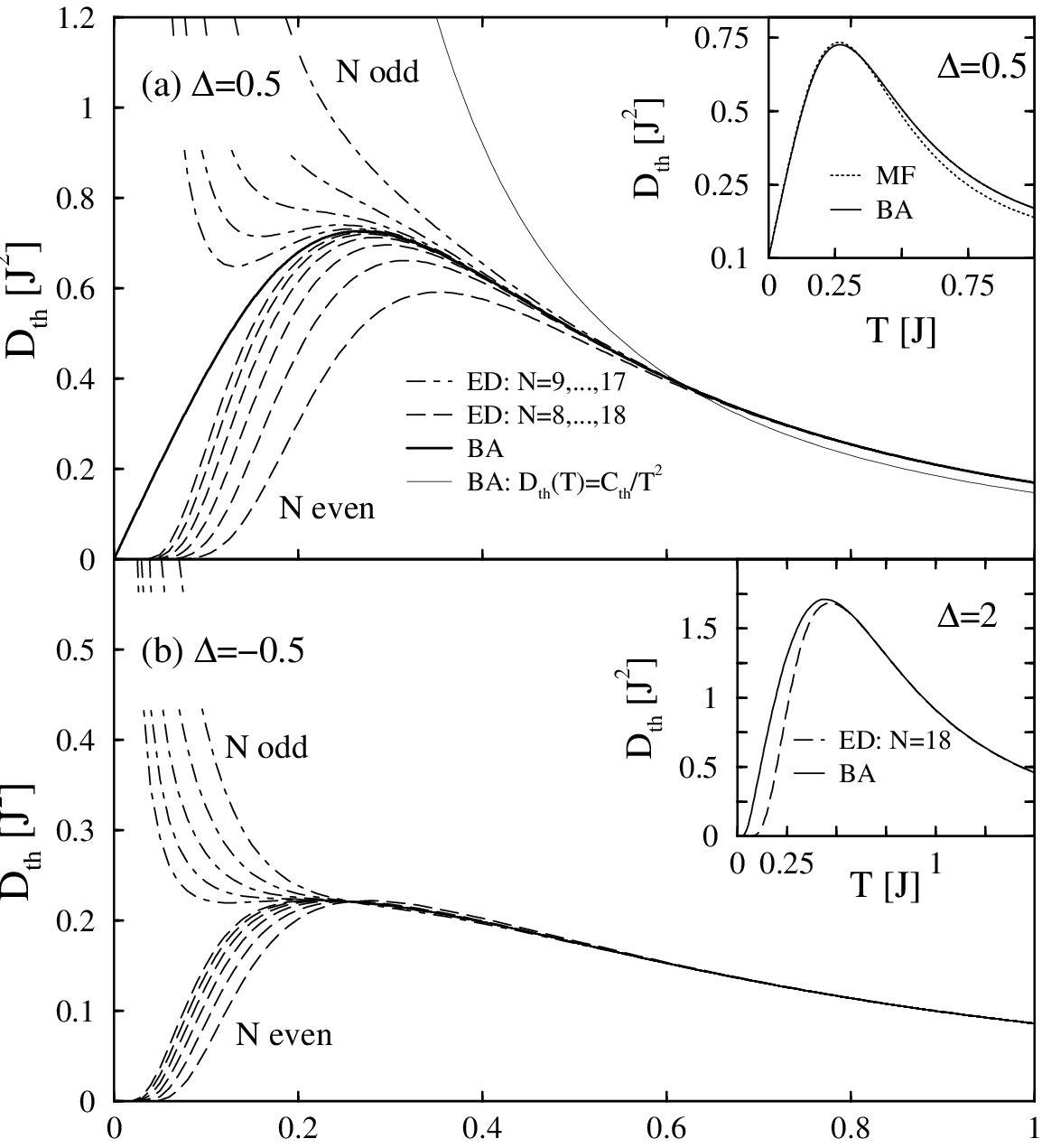,width=0.45\textwidth}}
\caption{
Thermal Drude weight of the $XXZ$ chain for: $\Delta=0.5$ (panel (a)),
$\Delta=-0.5$ (panel (b)) and $\Delta=2$ (inset in panel (b)). Dashed(dot-dashed) lines: ED data for even(odd) $N$.
Solid lines: Bethe ansatz  (BA) results from Refs.\ \onlinecite{kluemper02,kluemper03}. Thin, solid
line in (a): high-temperature limit $D_{\mathrm{th}}(T)\simeq C_{\mathrm{th}}/T^2$ from Ref.\
\onlinecite{kluemper02}.  Dotted line in the inset of (a): mean-field (MF) approximation.
\label{fig:1}}
\end{figure}
\begin{figure}[t]
\centerline{\epsfig{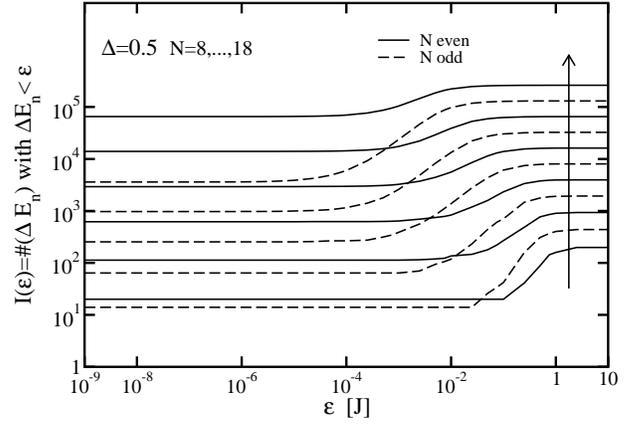}}
\caption{\label{fig:2} Distribution of level spacings in the spectrum of finite chains with $8\leq N\leq 18$ (bottom to top as indicated
by the arrow) for
$\Delta=0.5$ (solid lines: $N$ even, dashed lines: $N$ odd).
$\Delta E_n$ is the difference of adjacent energy levels in subspaces classified by total $S^z_{\mathrm{tot}}$
and momentum $k$. The number $I(\epsilon)$ of $\Delta E_n$ with $\Delta E_n< \epsilon$ summed over all subspaces
is plotted versus $\epsilon$ (see Eq.\ (\ref{eq:int-1})).
}
\end{figure}

\subsubsection{Thermal transport in the $XXZ$ model} 
\begin{figure}[t]
\centerline{\epsfig{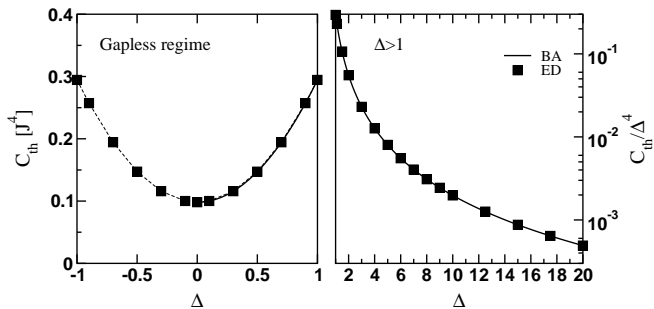}}
\caption{
High-temperature prefactor $C_{\mathrm{th}}$  for
the $XXZ$ model in the gapless (left panel) ($|\Delta|\leq 1$) and gapped, antiferromagnetic regime (right panel)
($\Delta>1$).
The numerical data (ED) are in perfect agreement with the exact
expression for $C_{\mathrm{th}}$ from the Bethe ansatz\cite{kluemper02} (BA).
\label{fig:4}}
\end{figure}

Numerical results for the thermal Drude weight $D_{\mathrm{th}}(N,T)$
have previously been presented for $\Delta=1$ in Refs.\ \onlinecite{gros02,hm02} and for $\Delta=-1;-2;10$
in Ref.\ \onlinecite{hm02}. We enhance on the latter study by adding numerical data for the gapless antiferromagnetic
and the ferromagnetic regime which is
shown in Fig.\ \ref{fig:1} (panel (a): $\Delta=0.5$,  (b) $\Delta=-0.5$).
Dashed(dot-dashed) lines display ED-data for $N$ even(odd). Since we are using periodic
boundary conditions, the twofold degeneracy of the ground state for chains with an odd
number of sites leads to a divergence of $D_{\mathrm{th}}(N,T)$ for $T\to 0$.
The Drude weight of systems with an even number of sites is characterized by an
exponential suppression
$D_{\mathrm{th}}(N,T)\sim e^{- \Delta_{\mathrm{FS}}/T}$  at low temperatures
where $\Delta_{\mathrm{FS}}$ is the finite-size gap of the system.\\
A striking feature of $D_{\mathrm{th}}(N,T)$ of the $XXZ$ model is
the rapid convergence with $N$ at high temperatures. The difference between
the numerical data for the largest system
investigated  for $\Delta=0.5$ (i.e., $N=18$)  and the Bethe ansatz curve\cite{kluemper02}
(solid line in Fig.\ \ref{fig:1} (a)) is smaller than $10^{-2}J^2$  for temperatures $T\gtrsim 0.23 J$.
Basically, the convergence of the numerical data is as good as in the case of other thermodynamic quantities
such as specific heat $C_{\mathrm{V}}$ and susceptibility $\chi$. We stress that 
the  relation $D_{\mathrm{th}}(T)\sim C_{\mathrm{V}}(T)$  only holds 
at high and low temperatures (see Ref.\ \onlinecite{kluemper02}).
\\
The antiferromagnetic, gapped regime has been discussed in Ref.\ \onlinecite{hm02}.
We mention that in the meantime Bethe ansatz computations\cite{kluemper03}
have been extended to $\Delta>1$. The numerical data
are in agreement with these results (see inset
of Fig.\ \ref{fig:1} (b): solid line
, BA; dashed line, ED for $N=18$ at $\Delta=2$).
\\
\indent
We continue by a discussion of the high-temperature prefactor
$C_{\mathrm{th}}$ defined in Eq.\ (\ref{eq:k6}). As  is evident from
Fig.\ \ref{fig:1} (a) and (b), the value of $C_{\mathrm{th}}:=\lim_{N\to\infty} C_{\mathrm{th}}(N)$
 can be numerically determined
 already from systems with a comparably small number of sites (e.g., $N=12$)
with very good accuracy.
Here we are interested in the $\Delta$-dependence of  $C_{\mathrm{th}}$
as shown in Fig.\ \ref{fig:4} (squares: ED).
Note the excellent agreement with the analytic expression from
the Bethe ansatz\cite{kluemper02} (solid line in Fig.\ \ref{fig:4})
\begin{equation}
C_{\mathrm{th}}(\Delta)= \frac{\pi \,J^{4}}{64}\left( 3+ \frac{\sin(3 \arccos(\Delta))}{\sin(\arccos(\Delta))} \right).
\label{eq:i1}
\end{equation}
For $\Delta=0$ (i.e., free fermions), we get $C_{\mathrm{th}}(\Delta=0)=\pi\,J^4/32 $
which can also be obtained from Eq.\ (\ref{eq:mf5}). \\
There are two main features of $C_{\mathrm{th}}(\Delta)$:
(i) $C_{\mathrm{th}}(\Delta)=C_{\mathrm{th}}(-\Delta)$
(see Fig.\ \ref{fig:4} (a))
and (ii) $\lim_{\Delta\to \infty}\lbrack C_{\mathrm{th}}(\Delta)/\Delta^4\rbrack=0$
(see Fig.\ \ref{fig:4} (b)). In Fig.\ \ref{fig:4} (b),
$C_{\mathrm{th}}$ is measured in units  of $\Delta$.\\
The first property can be seen from  the  following observations:
changing the sign of $\Delta$ is an antiunitary transformation and   essentially turns $H(-\Delta) = - H(\Delta)$
and $E_n(-\Delta) = -E_n(\Delta)$ while the eigenvectors remain unchanged (this follows from an additional
rotation by $\pi$ about the $z$ axis on all sites with  even site index).
This transformation leaves the  energy-current operator unchanged, i.e., $j_{\mathrm{th}}(-\Delta)=j_{\mathrm{th}}(\Delta)$.
From Eq.\ (\ref{eq:k7}), we can then conclude $C_{\mathrm{th}}(\Delta)=C_{\mathrm{th}}(-\Delta)$.\\
For  $J/\Delta\to 0$,  the Ising limit $H= \sum_l S_l^zS_{l+1}^z$ is approached where
the only possible excitations are local spin flips. Here,
no current can flow and, consistent with this notion, $\lbrack h_l,h_{l+1}\rbrack =0$
leading to $j_{\mathrm{th}}\equiv 0$ (see Eq.\ (\ref{eq:c3})) and $D_{\mathrm{th}}\equiv 0$ for $J/\Delta=0$. 
\subsubsection{Spin transport in the $XXZ$ model}
\label{sec:3b}
\begin{figure}[t]
\centerline{\epsfig{figure=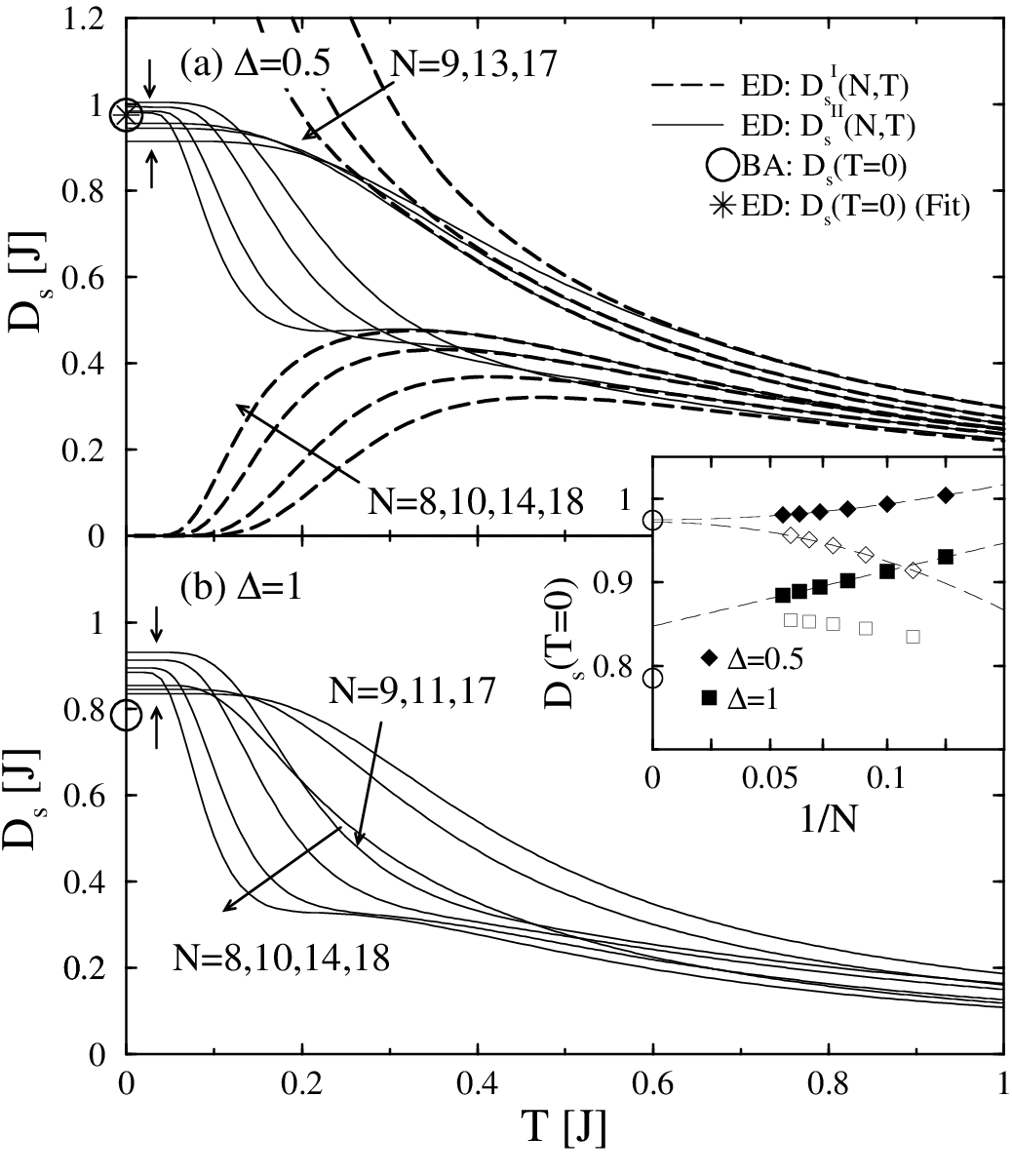,width=0.48\textwidth}}
\caption{Spin transport: Drude weight  for $\Delta=0.5$ (panel
(a)) and $\Delta=1$ (panel (b)). Dashed lines are numerical data
obtained for $D_{\mathrm{s}}^I(N,T)$ from Eq.\ (\ref{eq:k3}); solid
lines denote $D_{\mathrm{s}}^{II}(N,T)$ from Eq.\ (\ref{eq:k8}). Arrows
indicate increasing system size. 
Note that $D_{\mathrm{s}}^{II}(N,T)$ shows a  better convergence with $N$
at low temperatures compared to $D_{\mathrm{s}}^{I}(N,T)$.
The inset shows the Drude weight 
at $T=0$ for $\Delta=0.5$ (diamonds) and
$\Delta=1$ (squares) as a function of the inverse system size. The
dashed lines are fits applied to the subsets with even (solid
symbols) and odd (open symbols) $N$. Open circles at $T=0$
denote the exact values for $D_{\mathrm{s}}(T=0)$\cite{shastry90}.
\label{fig:5}}
\end{figure}

\begin{figure}[t]
\centerline{\epsfig{figure=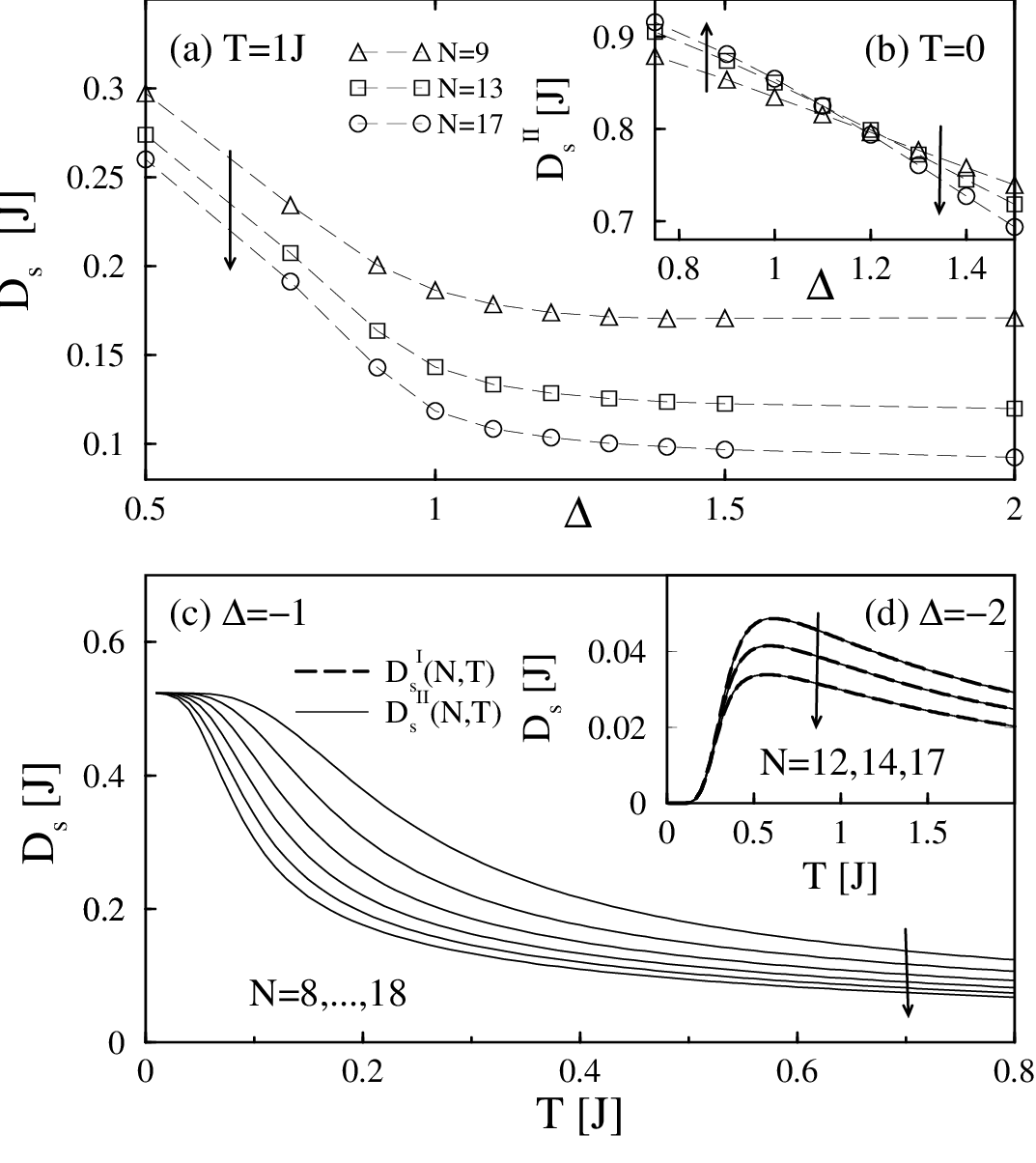,width=0.48\textwidth}}
\caption{Spin transport. Panels (a) and (b): Drude weight
$D_{\mathrm{s}}^{II}(N,T)$ as a function of $\Delta$. 
($N=9,13,17$; in (a): $T=1J$; (b):  $T=0$. Panel
(c): Drude weight $D_{\mathrm{s}}^{II}(N,T)$ for $\Delta=-1$ and
$8\leq N\leq 18$ ($N$ even, top to bottom). Panel (d): Drude
weight for $\Delta=-2$ and $N=12,14,17$ (top to bottom; dashed
lines: $D_{\mathrm{s}}^{I}(N,T)$ Eq.\ (\ref{eq:k3}), solid lines:
$D_{\mathrm{s}}^{II}(N,T)$ Eq.\ (\ref{eq:k8})). Arrows indicate
increasing system size.
\label{fig:6}}
\end{figure}

\begin{figure}[t]
\centerline{\epsfig{figure=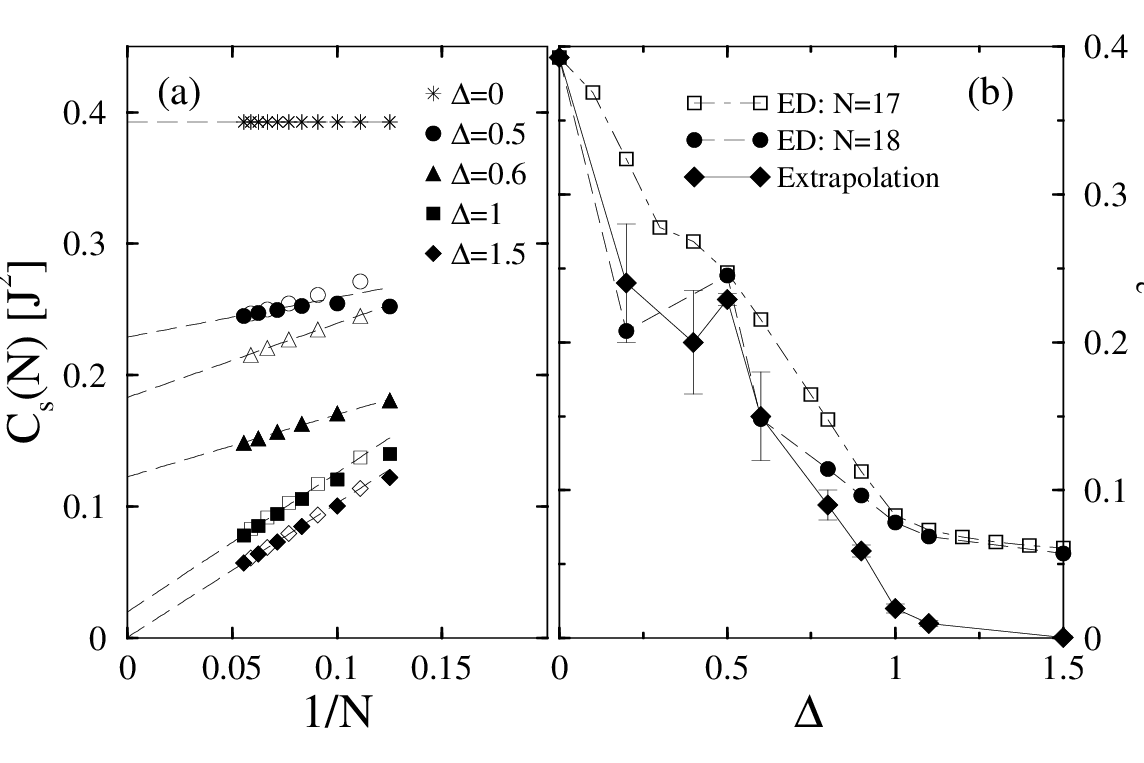,width=0.45\textwidth}}
\caption{
Spin transport. Panel (a): high-temperature prefactor $C_{{\mathrm{s}}}(N)$ versus $1/N$ for $\Delta=0,0.5,0.6,1,1.5$ 
(solid(open) symbols:
even(odd) $N$).
The dashed lines are fits according to Eq.\ (\ref{eq:int3}) (see text for details).
Panel (b): extrapolated high-temperature weight $C_{\mathrm{s}}(\Delta)$ as a function of the anisotropy $\Delta$ 
(diamonds) and the data for $N=17,18$ (squares, circles). The lines are guides to the eyes.
 \label{fig:7}}
\end{figure}

In the following, a  survey of the results for $D_{\mathrm{s}}$ for arbitrary values of the anisotropy will be given.
First, the gapless, antiferromagnetic regime ($0 < \Delta \leq 1$; Fig.\ \ref{fig:5}) will be discussed 
with a particular focus
on the isotropic chain ($\Delta=1$). Second, we comment on the finite-size data for $D_{\mathrm{s}}$
for the gapped, antiferromagnetic case ($\Delta>1$; Fig.\ \ref{fig:6} (a),\ (b)), and third, results for the ferromagnetic regime
will be shown ($\Delta<0$; Fig.\ \ref{fig:6} (c),\ (d)). Finally, the dependence of the high-temperature prefactor $C_{\mathrm{s}}$
on both anisotropy and system size is analyzed (Fig.\ \ref{fig:7}).\\
\indent
Numerical results  for  $D_{\mathrm{s}}^{I,II}(N,T)$ in the gapless,
antiferromagnetic  regime are shown in Fig.\ \ref{fig:5}
 for $\Delta=0.5$ (panel (a)) and  $\Delta=1$ (panel (b)).
Note that our  results for $D_{\mathrm{s}}^{I}(N,T)$  agree with
 the data for $\Delta=0.4$ (not shown in the figures) and $N\leq 14$ published in
Ref.\  \onlinecite{narozhny98} by Narozhny et al. \\
\indent
 First, we  concentrate on the case of  $\Delta=0.5$.
In panel (a), we compare data from Eqs.\ (\ref{eq:k3}) ($D_{\mathrm{s}}^{I}(N,T)$, dashed lines) and 
(\ref{eq:k8}) ($D_{\mathrm{s}}^{II}(N,T)$, solid lines) confirming that
these two expressions are equivalent at high temperatures ($T\gtrsim 0.5J$, depending on system size).

At low temperatures, $D_{\mathrm{s}}^{I}(N,T)$ shows much slower convergence with $N$ and essential
features of the temperature dependence are only present in the
data for $D_{\mathrm{s}}^{II}(N,T)$  which are
the finite value at $T=0$ and the vanishing slope of
$D_{\mathrm{s}}^{II}(N,T)$ for $T\to 0$. The latter observation
(i.e., $d\, D_{\mathrm{s}}^{II}(N,T)/d \, T=0$ at $T=0$) is consistent
with the Bethe ansatz by Kl\"umper and co-workers\cite{kluemper,glocke02}.
In general, the functional form of $D_{\mathrm{s}}(T)$ at low $T$
for $|\Delta|<1$ is
\begin{equation}
 D_{\mathrm{s}}(T)=D_{\mathrm{s}}(T=0)-\mathrm{const} \cdot T^{\alpha}
 \label{eq:int0}
 \end{equation}
where the exponent $\alpha$ depends on the anisotropy. In Ref.\ \onlinecite{zotos99},
the expression
$ \alpha=\frac{2}{\nu-1}$ was derived
(The integer $\nu$ parameterizes the anisotropy via
 $\Delta=\cos({\pi}/{\nu})$). Taking this result, Eq.\ (\ref{eq:int0}) would imply
$D_{\mathrm{s}}(T)=D_{\mathrm{s}}(T=0)-\mbox{const} \cdot T$ for $\Delta=0.5$ which is obviously not
consistent with our numerical data. For $|\Delta|<1$ and low $T$, the temperature dependence seems to be described by the expressions derived
by Fujimoto and Kawakami\cite{fujimoto03} which are compatible with QMC\cite{gros02q} and the ED presented here.\\
\indent
We have also determined
$D_{\mathrm{s}}^{II}(N,T=0)$ numerically by evaluating Eq.\
(\ref{eq:k8}) in the subspace containing the ground state (see inset of Fig.\ \ref{fig:5}). Using
Eq.\ (\ref{eq:k10}), one could go to larger systems than $N=18$
since only the curvature of the ground state is needed. However,
the main topic of this work is the Drude weight at finite
temperatures while the finite-size corrections for the $T=0$-Drude weight
have been computed in Ref.\ \onlinecite{laflorencie01}.\\
Data for  $D_{\mathrm{s}}^{II}(N,T=0)$ are plotted versus $1/N$ in
the inset of Fig.\ \ref{fig:5} for $\Delta=0.5$ (diamonds) and
$\Delta=1$ (squares). The data from finite systems with an even
number of sites form a monotonically decreasing sequence with $N$
at $T=0$  and small temperatures (see
Fig.\ \ref{fig:5} (a), $T\lesssim 0.2 J $) while the data for odd $N$  are a
monotonically increasing sequence. Thus, the results  for $D^{II}_{\mathrm{s}}(N,T)$ and even $N$
provide an upper bound and those from systems with odd $N$  a
lower bound for $D_{\mathrm{s}}(T)$ at low temperatures.\\ 
From Ref.\ \onlinecite{laflorencie01}, the 
leading finite-size corrections  at $T=0$ are available
\begin{equation}
D_{\mathrm{s}}^{II}(N,T=0) = D_{\mathrm{s}}(T=0) +
\frac{B}{N^{\mu}}+\dots\,
\label{eq:int2}
\end{equation}
with  $\mu = 2$ for $\Delta\lesssim 0.5$.
Performing  fits according to Eq.\ (\ref{eq:int2}) at $T=0$ separately for even(odd) $N$   
 results in $D_{\mathrm{s}}(T=0)= 0.9747(0.9717)J$
for $\Delta=0.5$
which is in very good agreement with the exact result\cite{shastry90} $D_{\mathrm{s}}(T=0)= 0.97428 J$.\\
\indent
At the isotropic point, i.e., $\Delta=1$, the curves display
similar features as for $\Delta=0.5$: (i) a vanishing slope for $T \to
0$, (ii) a monotonic decrease at high temperatures, and (iii)
$D_{\mathrm{s}}^{II}(2N,T)< D_{\mathrm{s}}^{II}(2N-2,T)$
($D_{\mathrm{s}}^{II}(2N+1,T)> D_{\mathrm{s}}^{II}(2N-1,T)$)
at low temperatures $T\lesssim 0.1J$.\\
 However, the finite-size data at the isotropic point and $T=0$ seem to follow $D_{\mathrm{s}}^{II}(N,T=0)=A+B/N$
in contrast to the case of $\Delta=0.5$  as can be seen in the inset of Fig.\ \ref{fig:5} (b).
For $A$, we find $A\approx 0.847 J$ which compares well with numerical results obtained by
the Lanczos method reported in Ref.\ \onlinecite{bonca94}.
Admittedly, a good approximation to the exact value of $D_{\mathrm{s}}(T=0)$ at $\Delta=1$
for $N\to \infty$ cannot be obtained from the numerical data since
the system sizes are far too small (see Fig.\ \ref{fig:5} (b)). In fact, from the work of Laflorencie et al.\cite{laflorencie01} it is known
that the relevant and leading  finite-size correction at $T=0$ and $\Delta = 1$ is  a logarithmic term. 
This is because
umklapp scattering is a marginally irrelevant perturbation in this case. Similar to the susceptibility
$\chi(T)$\cite{eggert94,laflorencie01},
$D_{\mathrm{s}}(T)$ is expected to show a sharp drop for $T\to 0$ accompanied with a diverging slope
at $T=0$\cite{kluemper} in the thermodynamic limit.\\
At sufficient large  temperatures, we believe that the numerical data for $\Delta=1$ presented in Fig.\ (\ref{fig:5}) (b)
give the correct
picture of the temperature dependence of the Drude weight. However,  Fujimoto and Kawakami\cite{fujimoto03}
have recently obtained
an analytic expression for $D_{\mathrm{s}}(T)$ in the low-energy limit with conformal field theory 
which is  compatible with our numerical data for  $|\Delta| < 1$ but not for $\Delta=1$. Here, Fujimoto and Kawakami
find $D_{\mathrm{s}}(T)<D_{\mathrm{s}}(T=0)$ while the data shown in  Fig.\ \ref{fig:5} (b) 
seemingly support the opposite relation.
Despite  this discrepancy, our  results do nevertheless support the notion of a {\it finite}
$D_{\mathrm{s}}(T>0)$ at $\Delta=1$. 
This is substantiated by the analysis of the high-temperature prefactor $C_{\mathrm{s}}(N)$
(see Eq.\ (\ref{eq:k7})) as we will discuss in detail below.\\
\indent
For larger $\Delta$ (i.e., $\Delta>1$), the monotonic increase
of $D_{\mathrm{s}}^{II}(N,T)$ at low temperatures for odd  $N$
changes to a monotonic decrease. This is illustrated in Fig.\
\ref{fig:6} (b) where $D_{\mathrm{s}}^{II}(N,T=0)$ is plotted versus
$\Delta$ for  $N=9,13,17$ (compare Refs.\ \onlinecite{laflorencie01,gu02}). The cross-over in the monotony occurs
at $\Delta\approx 1.2$, i.e., in the gapped regime. Since
$D_{\mathrm{s}}^{II}(N,T)\approx\mathrm{const}$ for small $T$, the
behavior at $T=0$ is characteristic for the low-temperature
regime. At larger temperatures, $D_{\mathrm{s}}^{II}(N,T)$ is a
monotonically decreasing function for both even and odd $N$ (see
panel (a) of Fig.\ \ref{fig:6} for odd $N$ and $T=1J$).\\
\indent
Regarding the  ferromagnetic regime (i.e., $\Delta<0$),
we concentrate on $\Delta=-1$ and $\Delta=-2$.
The results for $\Delta=-1$ plotted in Fig.\ \ref{fig:6} (c)
 indicate $ D_{\mathrm{s}}^{II}(N,T) \approx \mbox{const}$ at low $T$ with $D_{\mathrm{s}}(T=0)\approx 0.523(5) J$. However, since the low-energy spectrum for $\Delta=-1$
is of comparable complexity as for $\Delta=1$, one may expect nontrivial finite-size corrections
which could lead to
a different temperature dependence at low $T$. If the latter is true, then the system sizes are
too small to draw conclusions about the behavior of $D_{\mathrm{s}}^{II}(N,T)$ at very low $T$.\\
\indent The Drude weight in the
gapped, ferromagnetic regime is expected to 
show  a behavior  analogous to that of  $\Delta>1$. For instance, 
$D_{\mathrm{s}}^{II}(N,T)$ is monotonically decreasing with $N$ at
all temperatures irrespective of odd-even effects. Interestingly,
   $D_{\mathrm{s}}^{I}(N,T)$ and
 $D_{\mathrm{s}}^{II}(N,T)$ turn out to be
indistinguishable for $\Delta<-1$ and $N$ large enough which is
illustrated in   Fig.\ \ref{fig:6} (d). This plot shows
$D_{\mathrm{s}}^{I,II}(N,T)$ for $N=12,14,17$ at $\Delta=-2$   where
dashed lines denote $D_{\mathrm{s}}^I(N,T)$
from Eq.\ (\ref{eq:k3}) and solid lines stem from Eq.\  (\ref{eq:k8}) ($D_{\mathrm{s}}^{II}(N,T)$). \\
\indent
To conclude this section, we discuss the  high-temperature prefactor $C_{\mathrm{s}}$ both as a function of
anisotropy and system size.  In Fig.\ \ref{fig:7} (a), 
$C_{\mathrm{s}}(N)$  is shown versus $1/N$ for $\Delta=0,0.5,0.6,1,1.5$.
Results for the case of free fermions ($\Delta=0$) have been deduced 
 from Eq.\ (\ref{eq:mf6}) on finite systems
in the fermionic picture.
To fix the absolute values we note that $C_{\mathrm{s}} = \pi/8$ in our notation for the free-fermion case $\Delta=0$.\\
As is evident from Fig.\ \ref{fig:7} (a), $C_{\mathrm{s}}(N)$ roughly follows
\begin{equation}
C_{\mathrm{s}}(N)= a + \frac{b}{N}+ \dots\, .
\label{eq:int3}
\end{equation}
The same finite-size extrapolation has been used for
$D_{\mathrm{s}}^{I}(N,T) \cdot T$ in Ref.\ \onlinecite{narozhny98}
for even-numbered systems and $N\leq 14$ at $T=50J$. However, by
comparing the direct computation of $C_{\mathrm{s}}(N)$ from Eq.\
(\ref{eq:k7}) with fitting
\begin{equation}
D_{\mathrm{s}}^{I,II}(N,T)=\frac{C_{\mathrm{s}}(N)}{T}+\frac{C_1(N)}{T^2}+\dots
\label{eq:int4}
\end{equation}
to $D_{\mathrm{s}}^{I,II}(N,T)$, we find that, often, more than
one term in Eq.\ (\ref{eq:int4}) needs to be taken into account to
recover the result for $C_{\mathrm{s}}(N)$ from Eq.\ (\ref{eq:k7}). Consequently, it is
preferable to compute $C_{\mathrm{s}}(N)$ directly from  Eq.\ (\ref{eq:k7}).\\
The data can be well extrapolated according to  Eq.\ (\ref{eq:int3}) for $\Delta=0.5$ and $\Delta\geq 1$.
A subtlety arises from the strong differences between even and odd-numbered systems for intermediate
values of $\Delta$ such as $\Delta=0.6$ (see Fig.\ \ref{fig:7} (a)). In these cases (i.e., $\Delta=0.2,0.4,0.6,0.8,0.9$)
we have separately performed fits to the subsets with even and odd $N$. $C_{\mathrm{s}}$
is  then estimated by averaging
the results from the fits.
The large error bars for $\Delta=0.2,0.4,0.6,0.8,0.9$ are due to these large differences in the
extrapolated values of subsets with even and odd $N$.  
 \\
Following this procedure,  we obtain $C_{\mathrm{s}}= (0.15\pm 0.03)J^2$ for $\Delta=0.6$
while  the extrapolation of the even-numbered systems yields  $C_{\mathrm{s}}= (0.123\pm 0.001)J^2$. The latter value compares well
to the data published in Ref.\ \onlinecite{narozhny98} ($C_{\mathrm{s}}=(0.119\pm 0.004) J^2$).
The strong finite-size effects between even and odd numbered systems, however, indicate that
additional finite-size corrections apart from a simple $1/N$-term must become relevant for larger $N$
if the sequences of even and odd $N$ converge to the same value in the thermodynamic limit.
A difference in the extrapolated value for large $N$ of these two subsets does not seem to be plausible.\\
\indent
Note that it has been suggested in Refs.\ \onlinecite{naef98,long03}
to partly include spectral weight from low frequencies of $\mbox{Re}\,\sigma(\omega)$  to obtain $C_{\mathrm{s}}$ for all those
$\Delta$ which cannot be written as $\Delta=\cos(\pi/\nu),\nu$ integer. However, within such a procedure, a criterion is necessary 
to decide up to which frequency one should integrate $\mbox{Re}\,\sigma(\omega)$ and it goes obviously beyond the usual definition 
of the Drude weight via Kohn's formula (\ref{eq:k10}). In this paper, we prefer to confine our analysis to
 Eq.\ (\ref{eq:k8}), or (\ref{eq:k10}) respectively.\\
 \indent
  At the isotropic point $\Delta=1$, we have  fitted the numerical data for $C_{\mathrm{s}}(N)$ 
  using Eq.\ (\ref{eq:int3}) while varying the range of data
points ($N_{\mathrm{min}}\leq N \leq N_{\mathrm{max}}$). Stable fit results are obtained
for $11\leq N \leq N_{\mathrm{max}}$ with $N_{\mathrm{max}}>14$ and for
$N_{\mathrm{min}}\leq N \leq 18$ with $N_{\mathrm{min}}<14$.
We find $C_{\mathrm{s}}=(0.020\pm 0.006) J^2$\ indicating a finite Drude weight in the thermodynamic limit.
This result is slightly smaller than the data reported  in  Ref.\ \onlinecite{narozhny98}
($C_{\mathrm{s}}=(0.046\pm 0.005) J^2$ there for $N=6,8,\dots,14$)
but the latter is regained if we use our data for $N=6,\dots,14$. 
Since we observe
$C_{\mathrm{s}}(\Delta,N)=C_{\mathrm{s}}(-\Delta,N)$, this implies $D_{\mathrm{s}}(T>0)>0$ for the 
isotropic, ferromagnetic chain also.
 \\\indent
The results from the extrapolation are summarized in Fig.\ \ref{fig:7} (b) together with
our data  for $N=17$ and $N=18$.
The plot suggests that (i) $D_{\mathrm{s}}(T)>0$ for $\Delta=1$ and  (ii) $D_{\mathrm{s}}(T)=0$ for $\Delta \geq 1.5$ within our
numerical accuracy.
Respective conclusions can be drawn for $\Delta\leq -1$ since $C_{\mathrm{s}}(\Delta)=C_{\mathrm{s}}(-\Delta)$.
We  stress that
for intermediate Ising-like anisotropies (i.e., $1< \Delta < 1.5$) the system sizes may still be too small for
an unambiguous confirmation of the conjecture\cite{zotos96,naef98,peres99} $D_{\mathrm{s}}(\Delta>1)=0$. 
In particular, the possibility of a finite Drude weight in the gapped regime cannot be ruled out on the basis of the 
numerical data even though the Drude weight is zero at $T=0$. An example for such a scenario (i.e., $D_{\mathrm{s}}(T=0)=0$ but 
$D_{\mathrm{s}}(T>0)>0$) 
has been given in Ref.\ \onlinecite{kirchner99}.\\
Very recently, Long et al.\cite{long03} have applied a newly developed finite-temperature Lanczos method to 
compute $C_{\mathrm{s}}(N)$ for $N=24,26,28$. Compared to our data, their results strongly deviate from the 
fits  to Eq.\ (\ref{eq:int3}) for systems with $N\leq 18$ for all values of $\Delta$ presented in Fig.~\ref{fig:7}~(a). 

\section{Nonintegrable Models}
\label{sec:4}
In this section we address the issue of transport in nonintegrable models by means of
bosonization and exact diagonalization. As discussed in Sec.\ \ref{sec:cop}, both the spin-current and  heat-current 
operators
are not conserved in the presence of frustration and dimerization except for the case
of a dimerized $XY$ chain\cite{orignac02}.\\\indent
The original conjecture by Zotos and co-workers\cite{castella95,zotos96,castella96} stated that the Drude
weights are expected to vanish in nonintegrable models and we will argue in the following that 
our exact diagonalization study corroborates this statement.\\ 
In a first numerical work by Alvarez and Gros\cite{gros02} on 
thermal transport, the data obtained by complete and exact
diagonalization of systems with $N\leq 14$ have been interpreted in
favor of a nonzero Drude weight $D_{\mathrm{th}}(T>0)>0$ in the cases of spin ladders and
frustrated spin chains. However, we have argued\cite{hm02,hm03}
that this conclusion 
cannot be sustained for the case of gapped, frustrated chains if larger
systems of up to $18$ sites and additional values for the next-nearest-neighbor frustration $\alpha$ ($\alpha=0.35,0.5,1$) are
included in the finite-size analysis.\\
\indent
Additionally, several
authors have recently used analytic approaches to compute the Drude weight
$D_{\mathrm{th}}$\cite{orignac02,saito03} 
and the spin Drude weight $D_{\mathrm{s}}$\cite{fujimoto03} in the low-energy limit.
 If the effective model describes non-interacting particles  such as in Ref.\ \onlinecite{orignac02}
 in the case of the spin ladder,
  a finite Drude weight will naturally exist.
 However, to prove that the Drude weight is also finite in the corresponding lattice model 
 is  more difficult. 
The outcome of a low-energy description crucially depends on the 
effective models, i.e., one has to take care which operators are kept and which ones
can be omitted when passing from the lattice to the continuum limit. In particular, Rosch and Andrei\cite{rosch00} have  shown
that two independent incommensurate umklapp terms suffice to relax the spin 
 current in massless models. Since this result has not been fully 
appreciated by some of the aforementioned papers, we will discuss the line of reasoning of Ref.\
\onlinecite{rosch00} and apply  these ideas to the models which are of interest in this paper.
We will argue  that a vanishing Drude weight for both kinds of transport is expected 
in generic nonintegrable (massive) models also.
Thus, the first part of this section will  be devoted to the discussion of transport properties
in the continuum limit.\\
In Secs.\ \ref{sec:non2} and \ref{sec:non3}, we will complete our numerical investigation of both the
thermal and the spin Drude weight  of frustrated and dimerized spin systems with arbitrary values of $\alpha$ and
$\lambda$.
The main focus will be on the finite-size analysis of the high-temperature prefactor $C_{\mathrm{th[s]}}(N)$ (see Eq.\ (\ref{eq:k7})).
If not stated otherwise, $\Delta=1$.
\subsection{Bosonization}
\label{sec:non1}
The low-energy description of the systems studied below, i.e., dimerized chain and frustrated chains,
can be cast in the general form of a $U(1)$ scalar field theory, known as a  Luttinger
liquid (denoted by $H_{\mathrm{LL}}$), with a perturbation $g$ corresponding to a relevant operator $H_{\mathrm{rel}}$
plus all irrelevant operators $H_{\mathrm{irr}}$ allowed by the symmetries of the given 
problem\cite{schulz96} 
\begin{eqnarray}
H &=& H_{\mathrm{LL}}+H_{\mathrm{rel}}+H_{\mathrm{irr}},\label{eq:cft0} \\
 H_{\mathrm{LL}}&=& \int dx \left( v K (\partial_x \Theta)^2+ \frac{v}{K} (\partial_x \phi)^2 \right), \label{eq:cft1}\\
 H_{\mathrm{rel}}&=& g\, \int dx \cos(\alpha \phi)  \label{eq:cft2}.
\end{eqnarray}
$\phi=\phi(x,\tau)$ is a bosonic field in 1+1 dimensions and $\Theta$ is the dual field $\partial_x\Theta =(1/K)\partial_{\tau}\phi$.
$K$ is the Luttinger parameter and $v$ is the velocity.
General situations involving more than one relevant operator 
could also occur, but this does not change the discussion below. \\
The Hamiltonian, Eq.\ (\ref{eq:cft0}), with $H_{\mathrm{irr}}\equiv0$ corresponds either to a Luttinger liquid 
in a generic massless situation ($g =0$, e.g., the massless regime of the frustrated chain)
or to a sine-Gordon theory in the massive 
cases ($g \ne 0$, e.g., dimerized chain, massive regime of the frustrated chain). These descriptions provide in general the 
correct low-energy picture if one is interested in, e.g., the 
long-distance behavior of correlators, 
and usually, one can discard the irrelevant terms $H_{\mathrm{irr}}$
since they only contribute with subleading corrections.
\\
However, as was pointed out in Ref.\ \onlinecite{rosch00}, certain operators   have a crucial 
effect on transport properties and should therefore 
be taken into account to reproduce the correct low-frequency and 
low-temperature behavior even
if these operators are irrelevant in the renormalization group sense.
The main result of Ref.\ \onlinecite{rosch00} is that a certain class of incommensurate umklapp 
operators lead to the decay of all currents  and hence render all conductivities 
finite. The emerging picture is  that, except for very special 
circumstances which could happen in certain integrable models, one should 
expect a vanishing Drude weight and hence a finite conductivity.\\ 
It should be stressed that Ref.\ \onlinecite{rosch00} is devoted to 
massless cases, that is, to those situations  where no relevant operators are 
present ($g=0$ in Eq.\ (\ref{eq:cft2})). However, one can argue that the 
main ingredient in the proof,
that is, the violation of all conservation laws due to the presence 
of incommensurate umklapp operators, is independent of the scaling dimensions 
of the operators involved. Hence one could expect a similar picture in 
 the massive  case, while clearly the results for $\kappa[\sigma](\omega)$ will 
be quantitatively different. This conjecture is in full agreement with 
our numerical findings. \\
For pedagogical reasons let us briefly summarize the main results of Ref.\ \onlinecite{rosch00}: 
among the infinitely many irrelevant operators contained in  
$H_{\mathrm{irr}}$, those which could produce a decay of the currents are 
the incommensurate umklapp operators, which are, in the case of pure spin models,  generically of the form 
\begin{equation}
\int dx\,   {\cal O}_{n,m}(x)=\int dx\, g_{nm}\,\cos(\sqrt{2\pi} n\, \phi + k_{nm} x) \label{eq:cft3}.
 \end{equation} 
 $g_{nm}$ are coupling constants, $k_{nm}=2nk_F-mG$
where $k_F$ is the Fermi momentum, and $G$ is a reciprocal lattice vector. In a fermionic representation, $n$ is the number
of fermions which change chirality under the action of 
the operator ${\cal O}_{n,m}(x)$.\\
These operators do not 
modify the low-energy  expressions for   the energy and 
spin current. The same holds for the relevant operator\cite{saito03} in Eq.~(\ref{eq:cft2}). 
The currents take the form (see, e.g., Refs.~\onlinecite{hm02,rao01})
\begin{eqnarray}
j_{\mathrm{s}} &=&\frac{-vK}{\sqrt{2\pi}}  \int dx\, \partial_x \Theta, \label{eq:non12}
 \\
j_{\mathrm{th}} &=&v^2\int dx\, \partial_x \phi \partial_x\Theta .\label{eq:non13}
\end{eqnarray}
Comparing our notation with  Ref.\ \onlinecite{rosch00}, note that $j_{\mathrm{s}}\sim J_{\mathrm{0}}$ and $j_{\mathrm{th}}\sim P_T$. 
The spin current neither commutes with the  relevant operator nor with $H_{\mathrm{irr}}$ while for the thermal current,
$\lbrack H_{\mathrm{rel}} , j_{\mathrm{th}} \rbrack=0$
 but $\lbrack H_{\mathrm{irr}} , j_{\mathrm{th}} \rbrack\not= 0$.\\
  The key observation\cite{rosch00} is that in the presence of 
{\em one} such operator ${\cal{O}}_{n,m}(x)$, there is still a conserved current which can be written 
as a linear combination of the spin and  thermal current
\begin{equation}
j_{\mathrm{{conserved}}} = k_{nm} j_{\mathrm{s}}+  2n j_{\mathrm{th}} .
\label{eq:cft10}
\end{equation}
However, as soon as {\em more} than one of the operators ${\cal{O}}_{n,m}(x)$ are considered, 
no conservation law of the type (\ref{eq:cft10}) survives and hence the conductivity is expected to be finite. 
Since there is no reason {\it a priori} to exclude such incommensurate operators,
this seems to be the  generic situation. As incommensurate operators have  been
considered neither in Ref.\ \onlinecite{orignac02} nor in Ref.\ \onlinecite{saito03} it is not clear whether   their
results of a finite thermal Drude weight in the low-energy limit provide  a proof
of $D_{\mathrm{th}}(T>0)>0$
for   the respective nonintegrable lattice models.\\
\indent
Generally, even if the
current operator is not conserved, a nonzero Drude weight can be caused by the
existence of  (nontrivial) conserved quantities $\lbrace Q_l\rbrace$ with a finite
projection on the current  operator $j_{\mathrm{th[s]}}$ in the Liouville space (see, e.g., Refs.\
\onlinecite{brenig92,rosch00,forster,garst01}). 
 More precisely, the Drude weight is nonzero if
 \begin{equation}
 D_{\mathrm{th[s]}}(T)= \frac{\pi}{T^{2[1]}\,N}  ( j_{\mathrm{th[s]}} \,|\, {\cal{P}} j_{\mathrm{th[s]}} ) >0
\label{eq:cft20}
\end{equation}
 where ${\cal{P}}$ is the projection operator on {\em all} conserved quantities $\lbrace Q_l\rbrace$. 
 $( A \,|\, B)$ denotes 
 Mori's scalar product\cite{forster} in the space of operators
 \begin{equation}
 (A(t)\,|\, B)  =\frac{1}{\beta}\int\limits_{0}^{\beta}d\tau\, \langle A(t)^{\dagger} 
 B(i\tau)\rangle. 
 \end{equation}
Under certain circumstances, e.g., integrability, it is possible to construct an infinite set of $\lbrace Q_l\rbrace$ (see Ref.\ \onlinecite{zotos97} and
references  therein). Still, the evaluation of Eq.\ (\ref{eq:cft20})  is a nontrivial problem.\\    
In the literature\cite{zotos97,fujimoto03}, one often refers to a weaker condition than Eq.\ (\ref{eq:cft20}),
namely  Mazur's inequality\cite{zotos97,mazur69} where  a subset of the $\lbrace Q_l\rbrace $ or even only one 
operator $Q_i\in\lbrace Q_l\rbrace $
 is taken into account.
Therefore, Mazur's inequality provides a lower bound for the Drude weight.
For instance, the 
conservation law\cite{rosch00}, Eq.\ (\ref{eq:cft10}), would suffice to prove that $D_{\mathrm{s}}>0$ if 
no further incommensurate operators are considered.\\
Quite recently,   such an operator $Q_i$
has been found for charge transport in a Luttinger-liquid plus
interactions spoiling both the integrability and the conservation
of the spin current operator\cite{fujimoto03}, reformulating Zotos and co-workers's results from Ref.\ \onlinecite{zotos97} 
in the continuum limit.   The conserved quantity
can be readily identified as the thermal current operator (compare
Refs.\ \onlinecite{saito03,hm02}).  However, their proof of $( j_{\mathrm{s}} \, |\,Q_i)>0$
 assumes  particle-hole symmetry to be broken,
e.g., by the existence of a magnetic field (see also Refs.\ \onlinecite{zotos97,louis03}). 
Consequently, for the class of models
considered in Ref.\ \onlinecite{fujimoto03}, a nonzero Drude
weight can be inferred. Since  the 
incommensurate operators of Eq.\ (\ref{eq:cft3})
 are explicitly excluded in
Ref.\ \onlinecite{fujimoto03}, their result does not contradict
our numerical indications for  a vanishing Drude weight
$D_{\mathrm{s}}$ in nonintegrable spin-lattice models.
\subsection{Thermal transport in nonintegrable models}
\label{sec:non2}
\begin{figure}
\centerline{\epsfig{figure=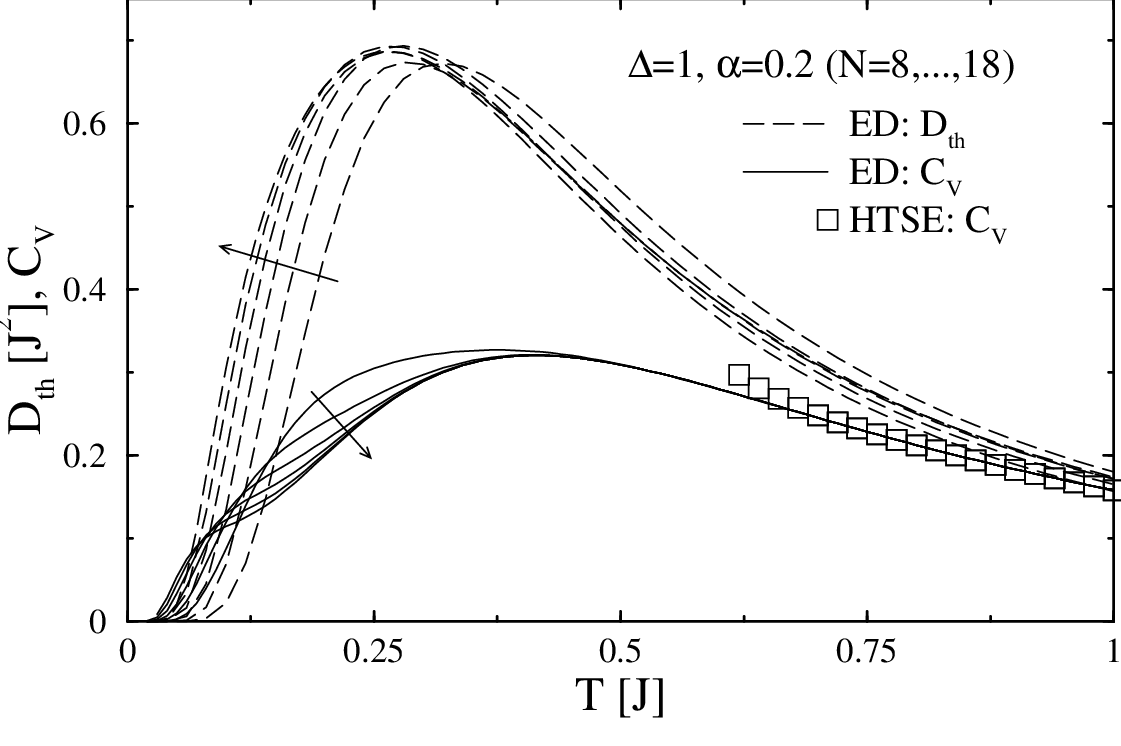,width=0.48\textwidth}}
\caption{
Thermal transport, frustrated chain: Drude weight $D_{\mathrm{th}}(N,T)$ (dashed lines) and specific heat $C_{\mathrm{V}}$ (solid lines) for $\alpha=0.2$
and $N=8,...,18$ sites (arrows indicate increasing system size). The plot includes data for $C_{V}$ from a high-temperature series expansion (HTSE)
reproduced from Ref.\ \onlinecite{buehler00} (note that only the bare high-temperature series up to order 10 in $J/T$ is shown. By means
of extrapolation schemes,
the HTSE can be extended to significantly lower temperatures, see Ref.\ \onlinecite{buehler00}).
\label{fig:8}}
\end{figure}
\indent
{\it Frustrated chain - }
In the thermodynamic limit, the low-energy spectrum of a frustrated chain with
$\alpha< \alpha_{\mathrm{crit}}\approx 0.241$\cite{nomura93}
is gapless and gapped for $\alpha >\alpha_{\mathrm{crit}}$.
The thermal Drude weight in the gapped regime of frustrated chains has been discussed in detail in Refs.\ 
\onlinecite{hm02,hm03} where we found 
clear indications of a vanishing Drude weight for $N\to \infty$.
Figure \ref{fig:8} shows the thermal Drude weight $D_{\mathrm{th}}(N,T)$ and the specific heat $C_{\mathrm{V}}$ for
$\alpha=0.2$
and $N=8,10,12,14,16,18$.  For chains of finite length, the data at low temperatures  are
dominated by the finite-size gap.
Hence the Drude weight and  specific heat are exponentially suppressed for small $T$.
While the specific heat  converges to the thermodynamic limit at temperatures $T\gtrsim 0.25 J$,
strong finite-size  effects are present in the data for the Drude weight at all temperatures.\\
\indent
 At low temperatures, $D_{\mathrm{th}}(N,T)$
monotonically increases with system size similar to the case of $\alpha=0.35$ (see Fig.\ 3 in Ref.\ \onlinecite{hm02}).
In Ref.\ \onlinecite{hm03}, we have argued that the notion of an increasing Drude weight at low temperatures does
not  support the conjecture\cite{gros02} of a finite $D_{\mathrm{th}}$ for $N\to \infty$
 for $\alpha=0.35$.
In fact, we have shown that a crossover temperature $T^*$ which we define  by $D_{\mathrm{th}}(N+2,T^*)
=D_{\mathrm{th}}(N,T^*)$  and even $N$ seems to extrapolate to zero  as a function of system size.  This implies that
the temperature range where one observes an increasing Drude weight with system size could vanish for
$N\to \infty$.
An analogous finite-size analysis of  $T^*$
for $\alpha=0.2$ (not shown in the figures) could also be interpreted in the same sense, i.e.,
  the temperature interval where $D_{\mathrm{th}}(N,T) < D_{\mathrm{th}}(N+2,T)$
tends to vanish for $N\to \infty$. To summarize the discussion of the low-temperature regime,
 we emphasize that the finite-size data should  not be used to speculate about the thermodynamic limit.\\
\begin{figure}
\centerline{\epsfig{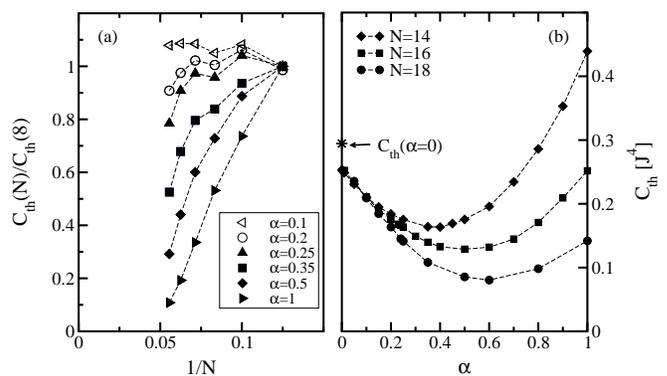}}
\caption{Thermal transport: high-temperature prefactor $C_{\mathrm{th}}(N)$ for frustrated chains.
Panel (a): $C_{\mathrm{th}}(N)/C_{\mathrm{th}}(8)$ versus $1/N$ for various values of $\alpha$
both in the gapless and gapped regime ($N=8,...,18$) (see also Ref.\ \onlinecite{hm03}). Panel (b): $C_{\mathrm{th}}(N)$ for
$N=14,16,18$ as a function of frustration $\alpha$. The arrow indicates the system size
independent value of $C_{\mathrm{th}}$ at $\alpha=0$.
\label{fig:9}}
\end{figure}
\indent 
In Fig.\ \ref{fig:9} (a), we present the high-temperature prefactor $C_{\mathrm{th}}(N)$
for several values of $\alpha$ both in the gapless ($\alpha=0.1,0.2$, open symbols)
and the gapped regime ($\alpha=0.25,  0.35,0.5,1$, solid symbols; the last three sets have already been shown in Ref.\
\onlinecite{hm03}). Finite systems with up to $N=18$ sites have been
analyzed.  While $C_{\mathrm{th}}(N)$ appears to be almost constant in the case of $\alpha=0.1$,
a substantial decrease with system size is observed for larger $\alpha$ and sufficiently large $N$ as it  is
especially obvious for $\alpha=1$. We also note that, 
regarding thermal transport, the data at the Majumdar-Ghosh point $\alpha=0.5$ (see Ref.\ \onlinecite{shastry81}
and references therein)
do not point to any peculiarities.\\
\indent
 In panel (b) of Fig.\ \ref{fig:9}, $C_{\mathrm{th}}(N)$
is plotted versus $\alpha$ for $N=14,16,18$. Starting at small $\alpha$, 
we observe that $C_{\mathrm{th}}(N)$ is discontinuous at $\alpha=0$ which will be
commented below.
The curve further decreases with $\alpha$ and exhibits a minimum at $\alpha\approx 0.4$ for $N=14$ and $\alpha\approx 0.5$
for $N=16$. The  position of the minimum seems to be further shifted towards larger $\alpha$ on growing $N$.
Further increasing  the next-nearest-neighbor interaction drives the system into the limit of
two decoupled chains each with $N/2$ sites and interchain interaction $\alpha$. Exactly for 
$J \to 0; J\alpha=\mathrm{const} $,
the current operator is again conserved. Consequently, one expects the Drude weight to increase for
large $\alpha $ at finite and fixed $N$. This feature is indeed found for $\alpha\gtrsim 0.5$, see Fig.\ \ref{fig:9} (b).
\\\indent
Fig.\ \ref{fig:9} (b) indicates a difference between the gapped and  gapless regimes: the decrease of $C_{\mathrm{th}}(N)$
with $N$ is weaker in the gapless regime. We suggest the following two   scenarios for further discussion:
(i) the Drude weight is nonzero in the gapless regime and zero in the gapped regime;
(ii) the Drude weight is zero for all $\alpha > 0$, but, depending on $\alpha$,
there is a characteristic system size $N(\alpha)$ with $C_{\mathrm{th}}(N)\approx\mbox{const}$ for $N<N(\alpha)$
and monotonically decreasing for $N> N(\alpha)$.\\
\begin{figure}[t]
\centerline{\epsfig{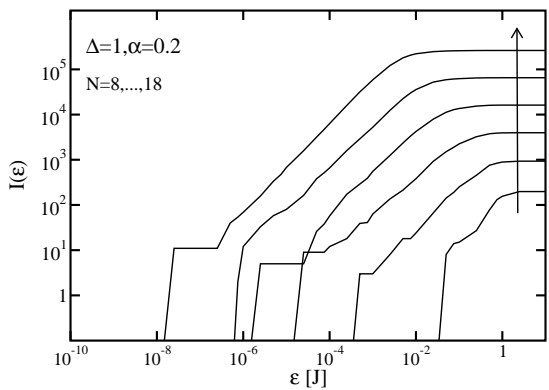}}
\caption{
Frustrated chain:
distribution of level spacings in the spectrum of finite chains with $8\leq N\leq 18$ (bottom to top
as indicated by the arrow) for
$\alpha=0.2$.
$\Delta E_n$ is the difference of adjacent energy levels in subspaces classified by total $S^z_{\mathrm{tot}}$
and momentum $k$. The number $I(\epsilon)$ of $\Delta E_n$ with $\Delta E_n< \epsilon$ summed over all subspaces is plotted
 versus $\epsilon$.
\label{fig:10}}
\end{figure}
\indent
The first interpretation might be plausible in view of the significant differences in the low-energy properties for
$\alpha<\alpha_{\mathrm{crit}}$ and $\alpha>\alpha_{\mathrm{crit}}$. However, since $C_{\mathrm{th}}$ is essentially
the Drude weight at {\it infinite} temperature where all states contribute with equal weight, 
it is not clear why low-energy features should play a crucial role for the finite-size scaling in the limit $\beta\to
0$.\\
 A second objection against the first scenario arises from the analysis of the level-spacing distribution
both in the gapped and gapless regime. Exploiting translational invariance and conservation of total $S^z_{\mathrm{tot}}$ 
already lifts all degeneracies on finite systems as  is obvious from Fig.\ \ref{fig:10}
showing the integrated level spacing distribution $I(\epsilon)$; see Eq.\ (\ref{eq:int-1}).
 The difference to the spectrum
of the integrable model (see Fig.\ \ref{fig:2}) is striking: while a large fraction of  states with  $\Delta E_n < 10^{-8}J$
is present for $\Delta=0.5, \alpha=0$, no such candidates for degenerate states appear in the case of $\alpha=0.2,\Delta=1$.
This feature is  characteristic  for $\alpha>0$ which,
in particular, supports the
conjecture that transport properties in the gapped and gapless regimes should not be different at high temperatures.
Exceptions are found for $\alpha=0.5,1$. At the Majumdar-Ghosh point, one degenerate state occurs if $N/2$ is even.
In the latter case (i.e.,
$\alpha=1$), there are degenerate states in the spectra of chains with $N=10,12,14,18$ which are, however, small
in number ($\approx 10$ for $N=18$). \\
\indent
Next, there is the discontinuity of $C_{\mathrm{th}}(N)$ at $\alpha=0$. A small, but finite frustration
(e.g., $\alpha=10^{-4},10^{-3}$) has the effect that  degeneracies are lifted while the values of
diagonal matrix elements $|\langle n \,|j_{\mathrm{s}}|\, n\rangle|^2$ are almost unaffected. 
This leads to the substantial difference between $C_{\mathrm{th}}$ at $\alpha=0$ compared to small, but 
finite $\alpha>0$.
Finally, we mention  that the fact of  $C_{\mathrm{th}}(N)\approx \mbox{const}$ for small $\alpha$ could be a
consequence of the proximity to the integrable point  $\alpha=0$.\\
\indent
In conclusion, our numerical data  indicate a vanishing thermal Drude weight for arbitrary values
of $\alpha$ at high temperatures. This result is difficult to
reconcile with  the recent findings\cite{orignac02,saito03} of a
nonzero Drude weight in the continuum limit as we have discussed above that a crossover
from a nonzero to a zero Drude weight as a function of
temperature is not likely.
\\\\

\begin{figure}
\centerline{\epsfig{figure=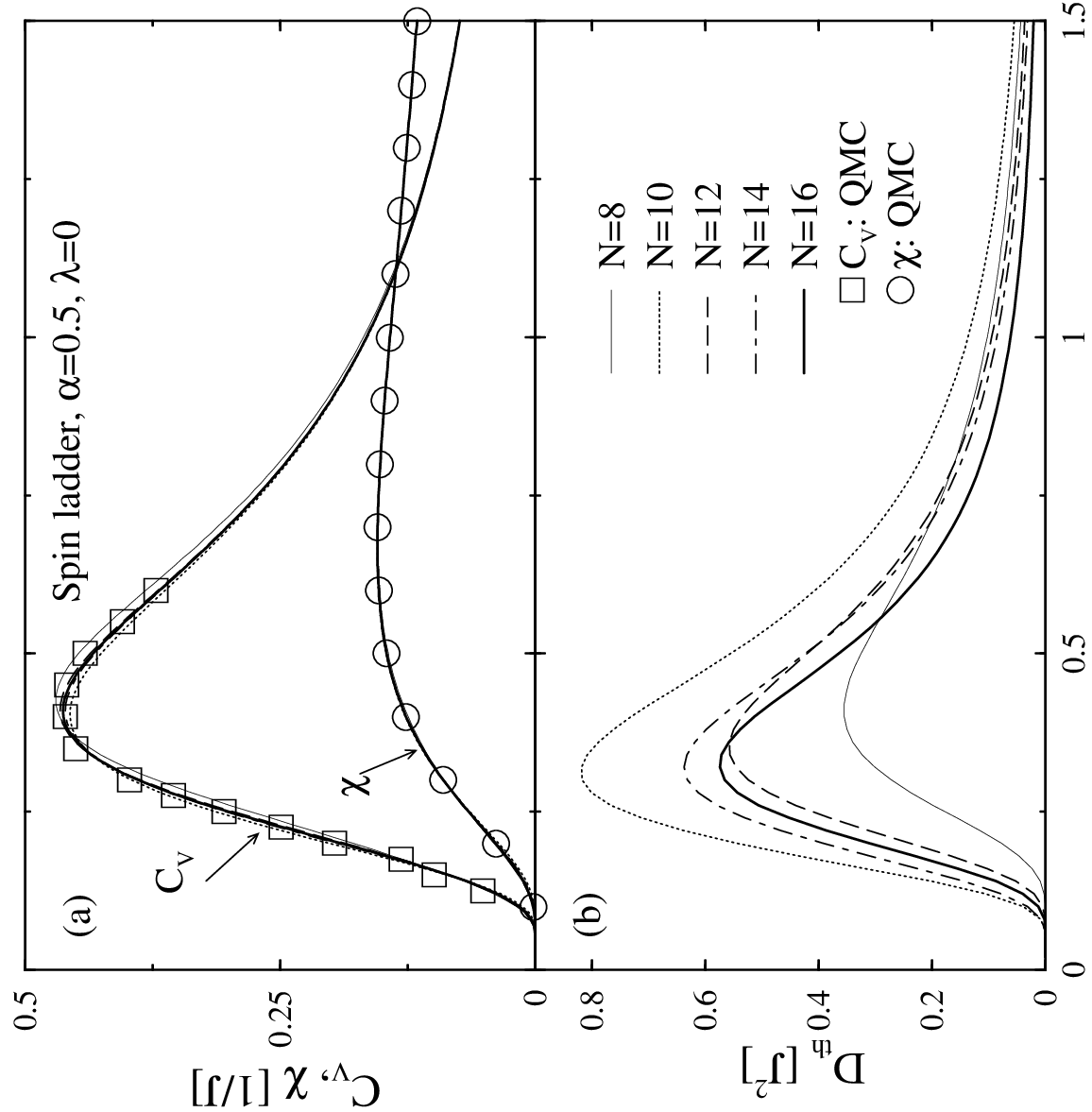,angle=-90,width=0.46\textwidth}}
\caption{
Thermal transport, spin ladder ($\alpha=0.5,\lambda=0$, $N=8,\dots,16$). Panel (a): specific heat $C_{V}$ and susceptibility $\chi$.
Panel (b): Drude weight
$D_{\mathrm{th}}(N,T)$.
ED for $D_{\mathrm{th}},C_V,\chi$: lines; QMC for $C_V$($\chi$): squares(circles) reproduced from
Ref.\ \onlinecite{gros02}(\onlinecite{johnston00}).
\label{fig:11}}
\end{figure}

\indent
{\it Spin ladder and dimerized chain - }
Now we turn to the cases  of  the dimerized chain and the spin ladder.
First, we discuss the numerical data for the thermal Drude weight $D_{\mathrm{th}}(N,T)$
taking the example of a spin ladder with $\alpha=J_{\parallel}/J_{\perp}=0.5, J_{\perp} =J$ for $N=8,10,12,14,16$.
Second, the results from a finite-size analysis of the high-temperature prefactor for both  spin ladders
and dimerized chains are presented. Finally, we comment on possible implications of our results for
the interpretation of recent experiments\cite{hess01}.\\
Due to the dimerization, the unit cell of our model is doubled restricting the maximal system size
to $N=16$ in our numerical computations at present.\\\indent
Regarding the level-spacing distribution, degeneracies are still present in the case of the spin ladder. For example,
there are   $\lesssim 10^2$ degenerate states for $N=16$ compared to $\gtrsim 10^3$ in the integrable case (see Fig.\ \ref{fig:2}).
 The spectra of dimerized chains show the same  features.\\
 \indent
The specific heat $C_{\mathrm{V}}$, the susceptibility $\chi$, and the thermal Drude weight $D_{\mathrm{th}}(N,T)$
are plotted versus $T$
 in Fig.\ \ref{fig:11} for $N=8,10,12,14,16$
and $\alpha=0.5,\lambda=0$ (panel (a): $C_{V}$, $\chi$, panel (b): $D_{\mathrm{th}}(N,T)$). The main characteristics are
the following:
(i) for the specific heat, finite-size effects are small and negligible
 for the susceptibility; (ii)
the data for $D_{\mathrm{th}}(N,T)$ display strong finite-size effects at all temperatures; (iii)
$D_{\mathrm{th}}(N,T)$ is monotonically decreasing at high temperatures for $N>8$ and $T\gtrsim 0.6 J$; (iv) the positions of the maxima
of the specific heat and the Drude weight are different;
(v) for $N/2$ even(odd), the data are monotonically increasing(decreasing) at low temperatures. The latter may be attributed
to the fact that $D_{\mathrm{th}}(N,T)$ is diverging for decoupled chains and odd $N$.\\
It should be stressed  that the  restricted number of system sizes analyzed here precludes 
any conclusions from the finite-size scaling at temperatures $T \lesssim 0.6 J$; in particular, since the $N$-dependence is 
nonmonotonic. Note that even a monotonic increase of $D_{\mathrm{th}}(N,T)$ with system size at low $T$
as observed in the case of frustrated chains with,  e.g., $\alpha=0.35$
does not unambigously point to a finite Drude weight 
(see the discussion of $D_{\mathrm{th}}(N,T)$ of frustrated chains and Ref.\ \onlinecite{hm03}).\\
\indent
While our numerical data for the specific heat are in quantitative agreement  with the results from Ref.\ \onlinecite{gros02}
for the same choice of parameters,
only qualitative  consistency is found for the thermal Drude weight regarding
points (ii)-(iv). Note that the Drude weight $D_{\mathrm{th}}$ in Ref.\ \onlinecite{gros02} is measured in units of
$J_{\parallel}\sim \alpha\, J$ instead of   $J_{\perp}=J$ used in this work. The specific heat, however, is dimensionless.
 Also, compared to point (v), the opposite monotony behavior
is observed at low temperatures in Ref.\ \onlinecite{gros02}.
One may speculate that this difference is due to the use of twisted boundary conditions  
and different 
definitions for the energy-current operator $j_{\mathrm{th}}$ in Ref.\ \onlinecite{gros02}.
\\
\begin{figure}
\centerline{\epsfig{figure=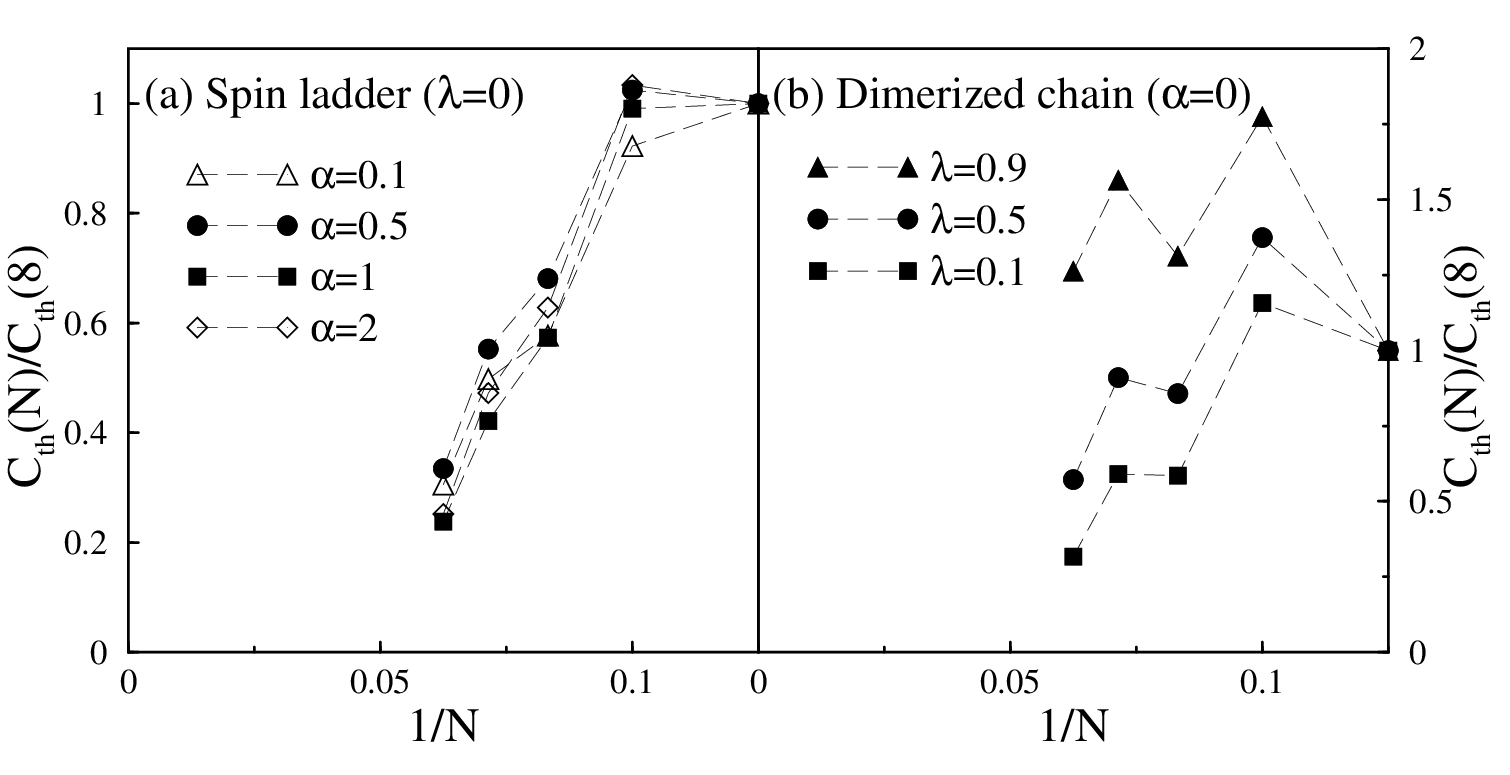,angle=-90,width=0.48\textwidth}}
\caption{
Thermal transport: $C_{\mathrm{th}}(N)$ for spin ladders (panel (a)) and  dimerized chains (panel (b)).
Data are shown for $N=8,\dots,16$ spins (i.e., ladders with $4,\dots,8$ rungs).
\label{fig:13}}
\end{figure}
\indent
The finite-size analysis of the high-temperature prefactor $C_{\mathrm{th}}(N)$ reveals a systematic
decrease with system size for $N> 8$  both for the case of the spin ladder and  dimerized chain as it is evident
from Fig.\ \ref{fig:13}. In particular, by normalizing the data on the respective values for $C_{\mathrm{th}}(N=8)$
the finite-size dependence of the data for the spin ladder (panel (a) in Fig.\ \ref{fig:13}) appears to be
almost
independent of the interchain coupling $\alpha=J_{\parallel}/J_{\perp}$ for the choice of parameters considered here
($\alpha=0.1,0.5,1,2$)  including  the isotropic ladder ($\alpha=1$).
Only the results for the dimerized chain with $\lambda=0.9$ show less evidence for
a vanishing of $C_{\mathrm{th}}$. This does, however, not question the conclusion of a vanishing thermal Drude
weight because $\lambda=0.9$ is still very close to the  homogeneous Heisenberg chain where $D_{\mathrm{th}}(T>0)$
is finite.\\
\indent
To summarize the finite-size analysis, one can conclude that the numerical data for $D_{\mathrm{th}}$
of spin ladders and
dimerized chains  indicate a vanishing Drude weight for $N\to \infty$.
In particular, this includes the isotropic spin ladder ($J_{\perp}=J_{\parallel}$;$\alpha=1$) which is of relevance 
because the
magnetic properties  of $\mbox{La}_{5}\mbox{Ca}_{9}\mbox{Cu}_{24}\mbox{O}_{41}$
are well described by $J_{\perp}\approx J_{\parallel}$\cite{matsuda00,windt01}.\\
\indent
Recently, first attempts have been made to extract magnetic mean free paths $l_{\mathrm{mag}}$ from the experimental data
for the magnetic part $\kappa_{\mathrm{mag}}$ of the thermal conductivity of $\mbox{La}_{5}\mbox{Ca}_9\mbox{Cu}_{24}\mbox{O}_{41}$.
Assuming that heat is  carried mainly by the elementary excitations of spin ladders (i.e.,
dispersive triplet modes)
 Hess et al.\cite{hess02} have used a relaxation time ansatz
 for the respective  kinetic equation reading
\begin{equation}
\kappa_{\mathrm{mag}} = \sum_k C_{\mathrm{V},k}\, v_k\, l_k .\label{eq:n1}
\end{equation}
Here, $C_{\mathrm{V},k}$ is the specific heat per $k$-space volume, $v_k$ is the triplet dispersion, and $l_k$
the momentum-dependent mean-free path.
This approach results in very large mean-free  paths $l_{\mathrm{mag}}$ of the order of 3000\AA\enspace 
at 100K corresponding to $\approx 770$ lattice constants. Following this work, Alvarez and Gros\cite{gros02}
have suggested a much smaller value
for $l_{\mathrm{mag}}$, namely, 176\AA\enspace$\approx$ 45 lattice constants  at $T=100K$.
They have attributed this significant difference to the fact that $D_{\mathrm{th}}(N,T)\not\sim C_{\mathrm{V}}$
as  is seen in the numerical data.
It is, however, straightforward to check that Eq.\ (\ref{eq:n1}) leads to $\kappa_{\mathrm{mag}}\not\sim C_{\mathrm{V}}$,
implying that this alone does not explain the different values found for $l_{\mathrm{mag}}$ in Refs.\ \onlinecite{hess01,gros02}.\\
The result for $l_{\mathrm{mag}}$ of Ref.\ \onlinecite{gros02}  is explicitly based on the
assumption of a {\it finite} Drude weight $D_{\mathrm{th}}$ for spin ladders which is
questionable in view of the detailed numerical results  presented in this paper.
\subsection{Spin transport in nonintegrable models}
\label{sec:non3}
In this last section, we give an  overview of our results for the
Drude weight for spin transport in nonintegrable models. The 
dependence of $D_{\mathrm{s}}$ on frustration and dimerization will be systematically 
discussed.
 Our analysis
of the Drude weight at finite temperatures includes next-nearest-neighbor
 interactions $\sum_l \vec{S}_{l}\cdot\vec{S}_{l+2}$
extending previous numerical studies of nonintegrable lattice
models\cite{narozhny98,gros02q,zotos96} where different kinds of Ising-like
 interactions ($ \sum_l S^z_{l} S^z_{l+i};\, i=2,3$) have been considered. 
 Note in this context that frustration cannot be  treated with QMC simulations due to the
 sign problem\cite{gros02q}. \\\\
\begin{figure}
\centerline{\epsfig{figure=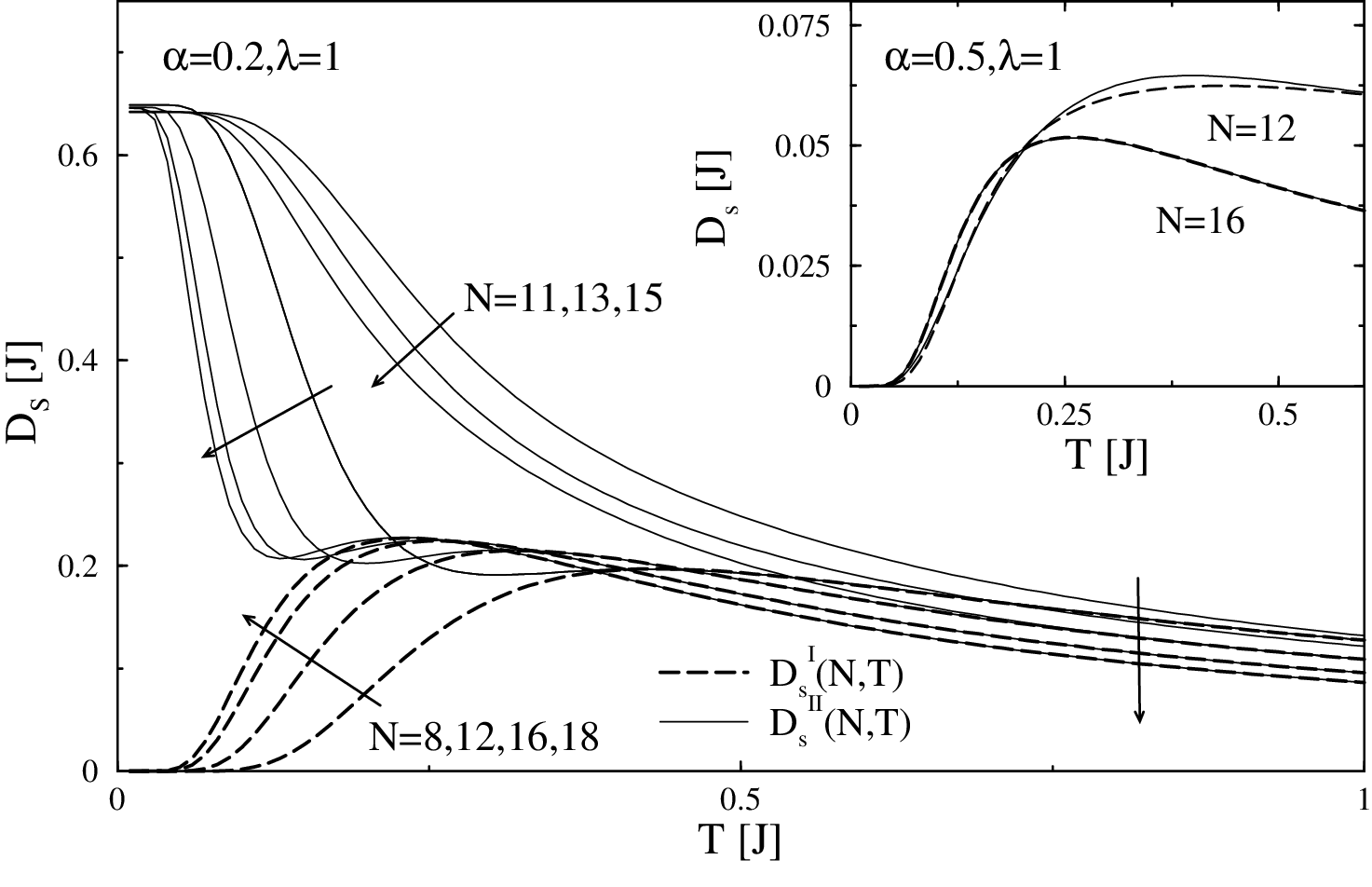,angle=-90,width=0.45\textwidth}}
\caption{Spin transport, frustrated  chain: Drude weight
$D_{\mathrm{s}}^{I,II}(N,T)$ for $N=8,11,12,13,15,16,18$ and $\alpha=0.2,
\lambda=1$  (dashed lines: $D_{\mathrm{s}}^I(N,T)$; solid lines: $D_{\mathrm{s}}^{II}(N,T)$).
Arrows indicate increasing systems size.
Inset:  Drude weight at the Majumdar-Ghosh point ($\alpha=0.5,\lambda=1$)  for $N=12,16$.
\label{fig:13a}}
\end{figure}
\indent
{\it Frustrated chain - }
Results for the Drude weight $D_{\mathrm{s}}^{I,II}(N,T)$ are shown in Fig.\ \ref{fig:13a} for $\alpha=0.2$  (main panel)
and $\alpha=0.5$ (inset). In the gapless regime ($\alpha<\alpha_{\mathrm{crit}}$), the finite-size data display features similar to the $XXZ$ model:
(i)  $D_{\mathrm{s}}^{II}(N,T=0)>0$;
(ii)  $D_{\mathrm{s}}^{II}(N,T)\approx \mbox{const}$ at small temperatures;
(iii) $D_{\mathrm{s}}^{I}(N,T)\simeq D_{\mathrm{s}}^{II}(N,T)$ at high temperatures, but significant deviations at 
low temperatures;
(iv)  a monotonic decrease with system size at high temperatures $T\gtrsim 0.5J$. Similar to the case of the thermal Drude weight of frustrated
chains, we observe that for even $N$, the monotonic decrease of $D_{\mathrm{s}}^{I,II}(N,T)$ with system size at high temperatures changes to 
an increase at lower $T$. However, the temperature, where this change in the monotony behavior  occurs, 
is strongly shifted to lower temperatures as $N$ grows; see Fig.~\ref{fig:13a}. This resembles the case of $D_{\mathrm{th}}(N,T)$ for $\alpha=0.35$ 
discussed in Ref.~\onlinecite{hm03} and we conclude that the numerical  data for $D_{\mathrm{s}}$ at low temperatures 
do not give unambiguous evidence
for a finite Drude weight.  
 \\
\indent
The observation of $D_{\mathrm{s}}^{II}(N,T=0)>0$ for small $\alpha$ has also been reported in Ref.\ \onlinecite{bonca94}. There, 
using the Lanczos method and truncation in the $S_{\mathrm{tot}}^z=0$ subspace, a nonzero Drude weight at $T=0$ has been found 
for $\alpha<0.43$ and $N=20$. 
To clarify whether $D_{\mathrm{s}}(T=0)>0$ survives in the thermodynamic limit, one should exploit Kohn's formula for $T=0$ 
using,  e.g.,
the Lanczos algorithm, which is, however, not the 
purpose of the present paper.\\\indent
Consistent with Ref.\ \onlinecite{bonca94}, we find $D_{\mathrm{s}}^{II}(N,T=0)=0$ in the case of $\alpha=0.5$ (see inset of Fig.\
\ref{fig:13a}). Notice that $D_{\mathrm{s}}^{I}(N,T)\simeq D_{\mathrm{s}}^{II}(N,T)$ at all temperatures for $N=16$.\\
We now turn to the question of a nonzero Drude weight in the thermodynamic limit by 
a finite-size analysis of the high-temperature prefactor $C_{\mathrm{s}}(N)$. $C_{\mathrm{s}}(N)$ 
is plotted versus $1/N$ in Fig.\ \ref{fig:13b} (a) for $\alpha=0.2,0.35,0.5,1$. 
First, $C_{\mathrm{s}}(N)$  monotonically decreases with system size (except for odd-even effects)
for all values of $\alpha$ presented here and exhibits a discontinuity at $\alpha=0$ analogous to
$C_{\mathrm{th}}(N)$ (see inset of Fig.\ \ref{fig:13b}). Second, the data for $\alpha=0.2$ may in principle 
be extrapolated to a finite value
in the thermodynamic limit (see, however, remarks below). The behavior for $\alpha>\alpha_{\mathrm{crit}}$ is similar to the 
case of the thermal Drude weight since $C_{\mathrm{s}}(N)$ decreases rapidly with $N$
and faster than $1/N$.\\\indent 
Another remarkable difference between the gapped and the gapless regime is revealed in Fig.\ \ref{fig:13b} (b) where we 
show $C_{\mathrm{s}}(N)$ as a function of frustration $\alpha$ for $N=14,16,18$. In contrast to the thermal 
Drude weight (see Fig.\ \ref{fig:9}), $C_{\mathrm{s}}(N)$ first grows with $\alpha$ and exhibits a maximum around
$\alpha\approx 0.2$. For larger $\alpha\gtrsim 0.7$, $C_{\mathrm{s}}(N)$ increases with $\alpha$ again analogous
to the case of the thermal Drude weight (see Fig.\ \ref{fig:9}). \\\indent
We interpret the data for $\alpha> \alpha_{\mathrm{crit}}$ in terms of a vanishing 
Drude weight for $N\to \infty$. 
For the gapless regime, the possibility for a nonzero  $C_{\mathrm{s}}$ cannot be 
ruled out by our  data although the absence of degeneracies (see Fig.\ \ref{fig:10}) supports the 
conclusion of $C_{\mathrm{s}}=0$ for all $\alpha>0$.
\\\\
\begin{figure}
\centerline{\epsfig{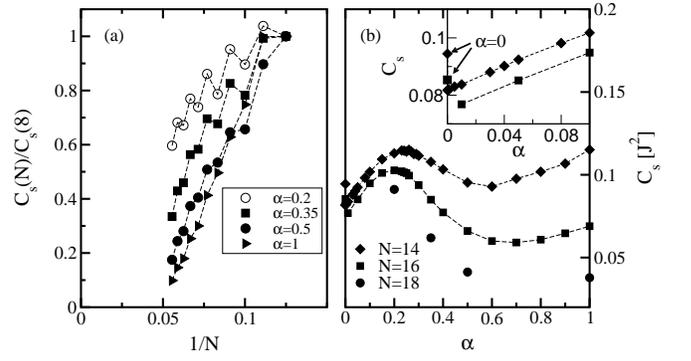}}
\caption{
Spin transport: high-temperature prefactor $C_{\mathrm{s}}(N)$ for frustrated chains.
Panel (a): $C_{\mathrm{s}}(N)/C_{\mathrm{s}}(8)$ versus $1/N$ for various values of $\alpha$
both in the gapless and gapped regime ($N=8,9,...,18$). Panel (b): $C_{\mathrm{s}}(N)$ for
$N=14,16,18$ as a function of frustration $\alpha$. The inset shows a blowup for small $\alpha$ and $N=14,16$.
\label{fig:13b}}
\end{figure}
\begin{figure}
\centerline{\epsfig{figure=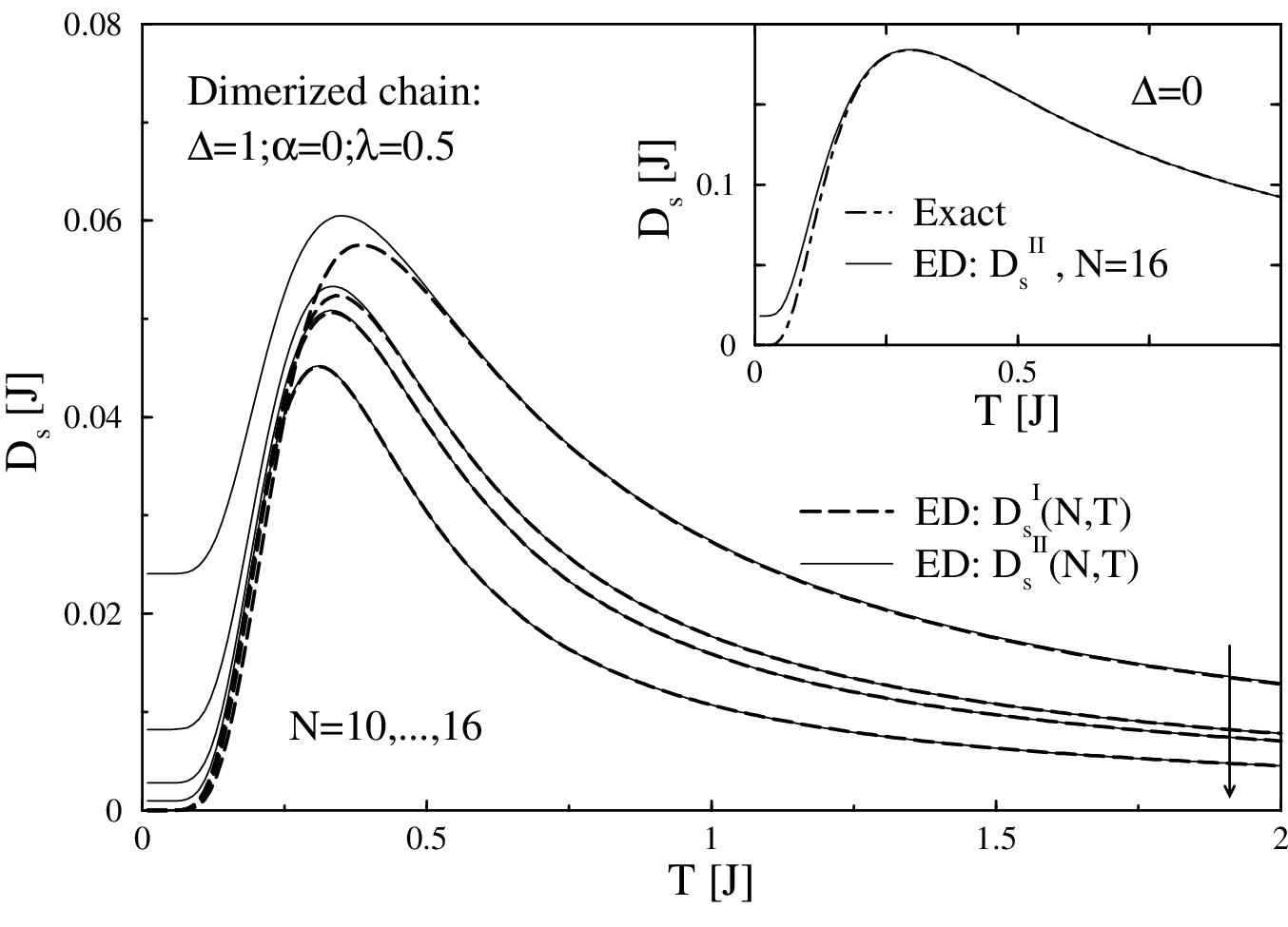,angle=-90,width=0.45\textwidth}}
\caption{Spin transport, dimerized chain: Drude weight
$D_{\mathrm{s}}^{I,II}(N,T)$ for $N=10,...,16$ (top to bottom as indicated by the arrow) and $\alpha=0,
\lambda=0.5$ (dashed lines: $D_{\mathrm{s}}^I(N,T)$; solid lines: $D_{\mathrm{s}}^{II}(N,T)$).  
In the inset, the exact result from Eq.\ (\ref{eq:n2})  (dot-dashed line) for the Drude weight of a dimerized 
$XY$ chain with $\Delta=0,\alpha=0,\lambda=0.5$ and numerical data (solid line) for $N=16$ sites are shown.   
\label{fig:14}}
\end{figure}

\indent
{\it Dimerized chain, spin ladder - }
While for the frustrated chain and the spin ladder the fermionized  Hamiltonian,  Eq.\ (\ref{eq:m5}),
contains interaction terms even at $\Delta=0$, 
the case of the  dimerized $XY$ model, ($\Delta=0,\lambda\not= 1,\alpha=0$) 
corresponds to a model of free, but massive fermions which can be solved exactly (see, e.g., Ref.\ \onlinecite{orignac02}).
We will start with a discussion of this limiting case where a finite Drude weight $D_{\mathrm{s}}$ exists.\\
The Hamiltonian in terms of spinless fermions (see Eq.\ (\ref{eq:m5})) reads
\begin{equation}
H^{XY} = \sum_l \frac{\lambda_l}{2} \, (c^{\dagger}_{l+1} c^{ }_{l} +\mbox{H.c.}) 
\label{eq:n3}
\end{equation}
($\lambda_l=\lambda$ for $l$ even and $\lambda_l=1$ otherwise). A straightforward computation 
diagonalizes  $H^{XY}$
(see, e.g., Ref.\ \onlinecite{orignac02} for details) 
\begin{equation}
H^{XY} = \sum_k \epsilon_{k} \, (a^{\dagger}_{k,+} a^{ }_{k,+}  - a^{\dagger}_{k,-}a^{ }_{k,-}),
\label{eq:n4}
\end{equation}
leading to two  modes with a gapped dispersion
$\epsilon_{k} = J\sqrt{(1-\lambda)^2/4 + \lambda \cos(k)}$.
Obviously, the spin-current operator 
$j_{\mathrm{s}}=\sum_k v_{k} (a^{\dagger}_{k,+} a^{ }_{k,+} -a^{\dagger}_{k,-}a^{ }_{k,-}) $
is conserved and  the Drude weight can be computed exactly
\begin{equation}
D_{\mathrm{s}}(T)= \frac{1}{4T}\int dk\, \frac{v_{k}^2}{\cosh^2\lbrack\epsilon_{k}/(2T)\rbrack}\,.
\label{eq:n2}
\end{equation}
 $v_{k}=\partial \epsilon_{k}/{\partial k}$ is the velocity. \\\indent
In Fig.\ \ref{fig:14}, the Drude weight on finite systems
 is plotted for a dimerized chain with
$\lambda=0.5$ and both $\Delta=0$ (inset) and $\Delta=1$ (main panel). 
In the former case, we see that the numerical data for $16$ sites agree with the exact expression for
$N\to\infty$,   Eq.\ (\ref{eq:n2}), 
at temperatures 
$T\gtrsim 0.2J$. The same holds for smaller system sizes (not shown in the figure) for high $T$. 
This is in contrast to the curves for $\Delta=1$ (main panel), where no convergence is observed at all temperatures.
\\
For $\Delta=1$, both  $D_{\mathrm{s}}^I(N,T)$  (dashed lines) and $D_{\mathrm{s}}^{II}(N,T)$
 (solid lines) have been evaluated proving their
equivalence at high temperatures. Small systems ($N=10,12$) still
exhibit a large nonzero value
of $D_{\mathrm{s}}^{II}(N,T=0)$ which rapidly decreases with system size.\\
\begin{figure}
\centerline{\epsfig{figure=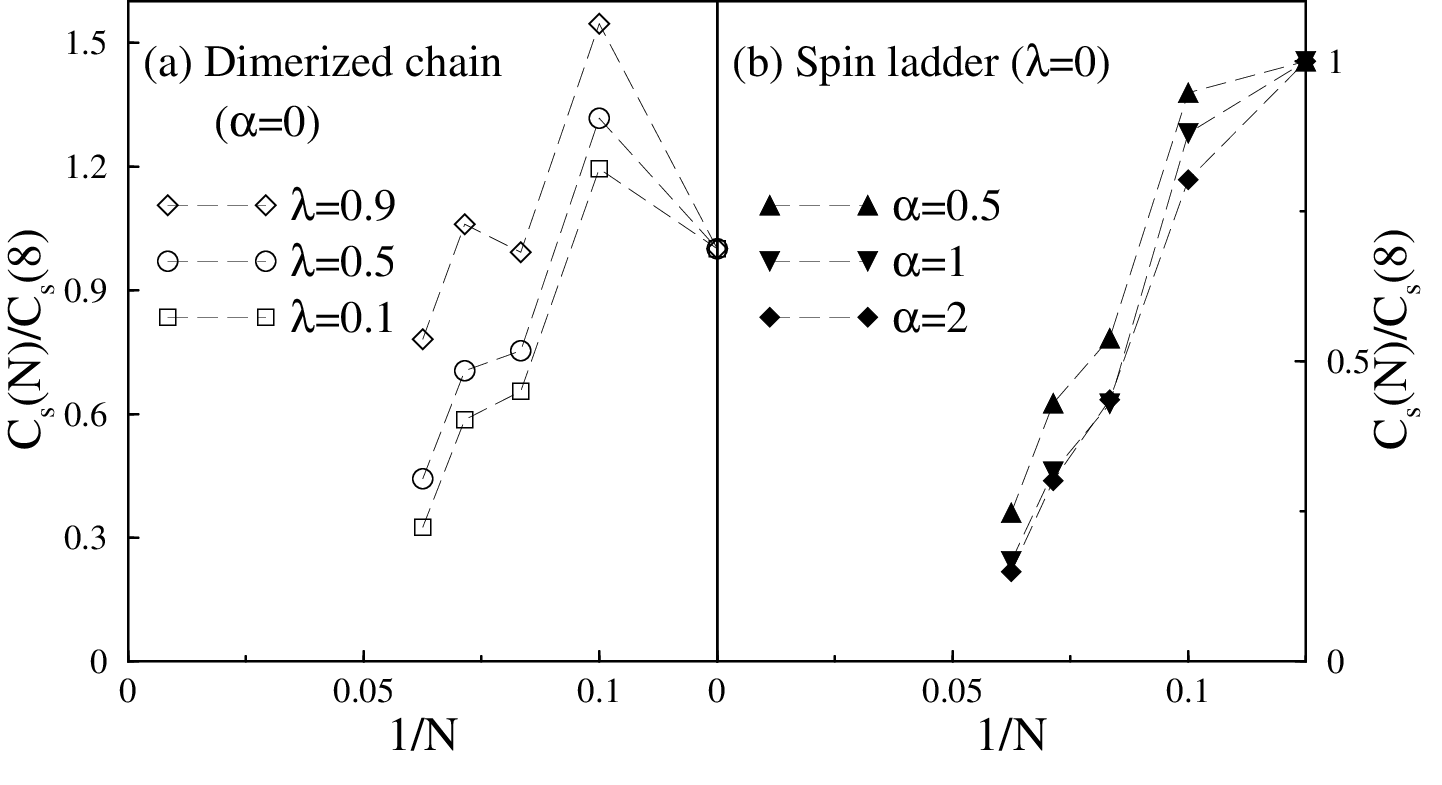,angle=-90,width=0.48\textwidth}}
\caption{
Spin transport: high-temperature prefactor $C_{\mathrm{s}}(N)$ of the Drude weight
 for the dimerized chain (panel (a)) and the spin ladder (panel (b)); both for $N\leq 16$.
\label{fig:15}}
\end{figure}
\indent
The data for the high-temperature prefactor $C_{\mathrm{s}}(N)$ are
collected in Figs.\ \ref{fig:15} (a) (dimerized chain) and (b)
(spin ladder). A substantial decrease of
$C_{\mathrm{s}}(N)$ with system size is observed for all choices
of parameters and $N>8$, indicating a vanishing Drude weight at high temperatures. \\\indent
In summary, our numerical data yield evidence for nonballistic spin transport
in nonintegrable spin models with a spin gap, i.e., $D_{\mathrm{s}}(T>0)=0$ in the
thermodynamic limit. This result  confirms  Zotos and coworker's
original conjecture\cite{zotos96} that transport in nonintegrable models 
should be normal. The system sizes investigated here may, however,  be
too small to clarify whether the Drude weight is zero or not in the gapless phase  
of the frustrated chain. \\\\
\section{Conclusion}
\label{sec:5}
In this paper, we have presented a detailed study of the Drude weights for thermal 
and spin transport in one-dimensional spin systems at finite temperatures. Let us focus here on the 
main results while more detailed summaries can be found in the preceding sections.\\
\indent
Thermal transport in the $XXZ$ model is generically dissipationless due to the conservation of
the energy current operator\cite{zotos97}. Our data for the thermal Drude weight are in excellent agreement 
with the Bethe ansatz\cite{kluemper02,kluemper03}. For spin transport in this model, we have presented numerical results for various 
values of the anisotropy. Our data confirm the observation\cite{naef98,zotos99,long03,zotos96,narozhny98,fujimoto03,gros02q} of a finite Drude weight
$D_{\mathrm{s}}$ 
in the gapless regime ($|\Delta|<1$).  We have discussed some so far unresolved issues:
first, the exact temperature dependence of the Drude  weight in the critical regime ($|\Delta|<1$)
and, second, the question of whether $D_{\mathrm{s}}(T>0)$ is finite for the isotropic chain ($\Delta=1$). 
Regarding the first point,  analytical\cite{fujimoto03} and numerical results\cite{gros02q}
are compatible with the finite-size data presented here.
Regarding the second issue, the exact diagonalization data of finite systems with $N\leq 18$ do favor a {\it finite} Drude weight
at $\Delta=1$. \\\indent
In the case of nonintegrable models (frustrated chain, dimerized chain, spin ladder), our main 
result is that the finite-size analysis of the ED data does not indicate a finite Drude weight in the thermodynamic limit 
either for thermal or spin transport, but  rather supports the conclusion that transport
in these systems is dissipative. While we have concentrated our numerical analysis on the finite-size scaling at  high temperatures,
this result is corroborated by bosonization in the low-temperature limit.\\
\indent {\bf Acknowledgments - }
This work was supported by the DFG, Schwer\-punkt\-programm 1073, and by
a DAAD-ANTORCHAS exchange program. It is a pleasure to thank J.\ V.\ Alvarez, N.\ Andrei,
B.\ B\"uchner, C.\ Gros, C.\ Hess, A. Kl\"umper, T.\ Lorenz, T. M. Rice,  and A. Rosch
for fruitful discussions.
We are indebted to A.\ Kl\"umper and K.\ Sakai for sending us their data.
We acknowledge support by the Rechenzentrum of the TU Braunschweig
where
parts of the numerical computations have been performed on a COMPAQ ES45.


\newpage
{\phantom{a}}
\newpage
In this Erratum, we correct five mistakes in our article Phys. Rev. B {\bf 68}, 134436 (2003).

First, there is a typographical error in Eq.~(10) ($E_n -E_m$ in the denominator of the second term). The equation should correctly read:
\begin{equation}
D_{\mathrm{s}}^{II}(N,T) = \frac{\pi}{N}\left\lbrack \langle  - \hat T\rangle -\frac{2}{Z} \sum_{m, n\atop E_m\not=E_n}
 e^{-\beta E_n}\frac{|\langle m |
j_{\mathrm{s}}|n\rangle|^2}{E_m-E_n} \right\rbrack.
\nonumber
\end{equation}
The correct expression was used in the numerical analysis.

Second, on page 5, we state that {\it Dimerization and frustration which spoil the integrability of the $XXZ$ model also lead to
$\lbrack H,j_{\mathrm{th[s]}}\rbrack\not= 0$ (except for $\Delta=\alpha=0$).}
This incorrectly implies that $\lbrack H, j_s \rbrack =0$ for the dimerized spin-1/2 XY chain. The spin current of the dimerized spin-1/2
XY chain is not conserved since it incorporates inter-band transitions.
Yet, the presence of a trivial Drude weight for the free dimerized chain ($\Delta=\alpha=0$) remains valid simply by virtue of Eqs.~(45) and (50) because the intraband current is  conserved.
See also the inset of Fig.~15 of the original paper, where numerical data are compared to Eq.~(50), showing agreement.
Our quantitative results for the spin Drude weight of this model are not affected.

Third, on page 15 below Eq.~(49), we give an expression for a spin-current operator of the dimerized spin-1/2 XY chain that is conserved. In the sentence "Obviously, the spin current
$j_s = \sum_k v_k (a^\dagger_{k,+} a_{k,+} - a^\dagger_{k,-} a_{k,-} )$ is conserved \dots ", the symbol $j_s$ needs to be replaced
by $j_s^{\rm diag}$, denoting the conserveed spin current that accounts only for intraband transitions, to distinguish it from the
spin-current operator derived in Eq.~(24).

Fourth, the numerical data for the spin Drude weight $D_s^I$ for $\Delta=0.5$ and $N=8$ presented in Fig.~5(a) of the original paper
has a small error due to neglecting  accidental degeneracies on this system size.
Figure~\ref{fig1} of this Erratum shows Fig.~5 of the original publication including the correct data for  $D_s^{I}$ computed for $N=8$ in panel (a). This error does not affect the extrapolation of the infinite-temperature Drude weight shown in the inset of Fig.~5 of the original paper.

Fifth, there is a typographical error in the dispersion relation of a dimerized spin-1/2 $XX$ chain quoted on page 15 below Eq.~(49)
(missing power of two for the cosine).
The correct expression is:
$$
\epsilon_{k} = J\sqrt{(1-\lambda)^2/4 + \lambda \cos^2(k)} \,.
$$
The correct expression was used in the calculation of the spin Drude weight from Eq.~(50) of the original paper.

\vspace{1cm}
All these corrections do not affect the discussion and conclusions in the original paper.
\vspace{1cm}

We thank J. Hauschild and C. Karrasch for bringing these errors to our attention.\\

\begin{figure}[b]
\centering
\includegraphics[width = 0.43\textwidth]{figure5_new.eps}
\caption{(Color online)
This figure replaces Fig. 5 of the original publication and includes the corrected data for  $D_s^{I}$ computed for $N=8$.
}
\label{fig1}
\end{figure}
\end{document}